\documentclass{aa} 

\usepackage{graphicx} 

\newcommand{\aFe}{\ensuremath{\alpha/{\rm Fe}}}
\newcommand{\Hb}{\ensuremath{{\rm H}\beta}}
\newcommand{\Mgb}{\ensuremath{{\rm Mg}\, b}}
\newcommand{\Fe}{\ensuremath{\langle {\rm Fe}\rangle}}

\newcommand{\ZH}{\ensuremath{Z/{\rm H}}}
\newcommand{\MgFe}{\ensuremath{[{\rm MgFe}]^{\prime}}}

\begin{document}

\title{Spatially resolved spectroscopy of Coma cluster early -- type 
galaxies: III. The stellar population gradients
%\thanks{} 
}
\author{D.\,Mehlert\inst{1} 
\and D.\,Thomas\inst{2} 
\and R.\,P.\,Saglia\inst{3} 
\and R.\,Bender\inst{2,3} 
\and G.\,Wegner\inst{4} 
} 

\offprints{D.\,Mehlert,\\
\email{dmehlert@lsw.uni-heidelberg.de}} 

\institute{Landessternwarte Heidelberg, K\"onigstuhl, 
  D-69117 Heidelberg, Germany 
\and Max-Planck-Institut f\"ur extraterrestrische Physik,
  Giessenbachstra\ss e, D-85748 Garching, Germany
\and Universit\"ats-Sternwarte M\"unchen, Scheinerstra{\ss}e~1, 
  D-81679 M\"unchen, Germany 
\and{Department of Physics and Astronomy, 6127 Wilder Laboratory, 
  Dartmouth College, Hanover, NH 03755-3528, USA}
}

\date{Received 7.3.2003; accepted 10.6.2003} 

\authorrunning{Mehlert et al.} 
%\titlerunning{Spectroscopy of Coma Early-Type Galaxies:
%   III. The stellar population gradients}
\titlerunning{Spectroscopy of Coma Early-type Galaxies: III. Gradients}

\abstract{Based on Paper I of this series (Mehlert et al. 2000), we
derive central values and logarithmic gradients for the \Hb, Mg and Fe
indices of 35 early-type galaxies in the Coma cluster. We find that
pure elliptical galaxies have on average slightly higher velocity
dispersions, lower \Hb, and higher metallic line-strengths than
galaxies with disks (S0).  The latter form two families, one
comparable to the ellipticals and a second one with significantly
higher \Hb, and weaker metallic lines.  Our measured logarithmic
gradients within the effective radius are
$\langle\triangle\Mgb\rangle\approx -0.037$, $\langle\triangle
\Fe\rangle\approx -0.029$, $\langle\triangle\Hb\rangle\approx +0.017$
and $\langle\triangle\sigma\rangle\approx -0.063$.  The gradients
strongly correlate with the gradients of $\sigma$, but only weakly
with the central index values and galaxy velocity dispersion.\\ Using
stellar population models with variable element abundance ratios from
Thomas, Maraston \& Bender (2003a) we derive average ages,
metallicities and [\aFe] ratios in the center and at the effective
radius.
%%%new
%{\bf 
We find that the \aFe\ ratio correlates with velocity dispersion
and drives 30\% of the Mg-$\sigma$ relation, the remaining 70\% being
caused by metallicity variations.  
%}
%%%
We confirm previous findings that part of the lenticular galaxies in
the Coma cluster host very young ($\sim 2$ Gyr) stellar populations,
hence must have experienced relatively recent star formation
episodes. Again in accordance with previous work we derive negative
metallicity gradients ($\sim -0.16$ dex per decade) that are
significantly flatter than what is expected from gaseous monolithic
collapse models, pointing to the importance of mergers in the galaxy
formation history. Moreover, the metallicity gradients correlate with
the velocity dispersion gradients, confirming empirically earlier
suggestions that the metallicity gradient in ellipticals is produced
by the local potential well.  The gradients in age are negligible,
implying that no significant residual star formation has occurred
either in the center or in the outer parts of the galaxies, and that
the stellar populations at different radii must have formed at a
common epoch.  For the first time we derive the gradients of the \aFe\
ratio and find them very small on the mean. Hence, \aFe\ enhancement
is not restricted to galaxy centers but it is a global phenomenon.
Our results imply that the Mg-$\sigma$ local relation inside a galaxy,
unlike the global Mg-$\sigma$ relation, must be primarily driven by
metallicity variations alone.  Finally we note that none of the
stellar population parameters or their gradients depend on the density
profile of the Coma cluster, even though it spans 3 dex in density.
}
\maketitle
%\markboth{Mehlert et al.: Major axis indices and gradients}{} 

\section{Introduction}
\label{Introduction}

This is the third paper of a series studying the dynamics and the
stellar populations of a sample of 35 early-type galaxies in the Coma
cluster, the richest of the local universe and therefore an ideal
place to test the theories of galaxy formation as a function of
environmental density. In Mehlert et al. (2000, hereafter Paper I) we
presented the photometry and the long-slit spectroscopy along the
major axis of the galaxies. In Wegner et al. (2002) 
this dataset is complemented by long-slit spectroscopy along the
minor axis and parallel to the major axis for 10 objects of the
sample. Here we focus on the radial profiles of the Mg, Fe and
H$\beta$ indices derived in Paper I, exploiting newly developed
stellar population models (Thomas, Maraston \& Bender 2003a, hereafter
TMB) to study the metallicity, the age and the \aFe\ distributions
inside the galaxies and relate these findings to the global relations
among the galaxies.

Recently, a number of papers have investigated the constraints on the
ages, metallicities and element abundances of Coma cluster galaxies
coming from (central) colors and spectroscopic (Lick) index
measurements and the possible influence of the environmental density.
J\o rgensen (1999) studies 115 E and S0 galaxies in the central region
of the cluster, suggesting that metallicities are strongly
anti-correlated with the mean ages of the galaxies.  Poggianti et
al. (2001a) analyze the spectra of 52 early-type Coma galaxies and
find that more than 40\% of the S0s are found to have undergone star
formation in their central regions during the last 5 Gyr, while such
activity seems absent in ellipticals (see also Trager et al.\
2003). Poggianti et al. (2001b) extend this study to a sample of 257
galaxies with no emission lines, spanning a wide magnitude range. They
find that age and metallicity are anti-correlated in any given
luminosity bin and that in the central regions of the cluster a large
fraction of galaxies at any luminosity shows no evidence for star
formation occurring in their central regions at redshift $z<2$. Carter
et al. (2002) focus on the possible environmental dependence of the
indices, finding that after allowing for the correlation with
magnitude, galaxies near the core of the cluster have stronger Mg$_2$,
while the $\langle$Fe$\rangle$ and H$\beta$ values show a much weaker
sensitivity to the cluster distance. In contrast, Kuntschner et
al. (2002) study a sample of ``field'' E/S0 galaxies in low-density
environments, reaching the conclusion that these objects are younger
than E/S0s in clusters (by $\approx 2-3$ Gyr) and more metal-rich (by
$\approx 0.2$ dex). They confirm that an anti-correlation of age and
metallicity is responsible for maintaining the zero-point of the
Mg-$\sigma$ relation (see also Kuntschner et al. 2001, pointing out
the role of correlated errors in generating this anti-correlation).

Similar to Terlevich \& Forbes (2002), these studies break the
metallicity-age degeneracy by combining pairs of indices with nearly
orthogonal dependences on metallicity and age, and are based on simple
stellar population models (Worthey 1994, Vazdekis et al. 1996). They
take into account the overabundance of the $\alpha$--elements with
ad-hoc assumptions (J\o rgensen 1999) or simple, approximated recipes
(Trager et al. 2000) and might be severely affected by systematics
induced by complex metallicity distributions (Maraston \& Thomas
2000). However, they provide a way to test the low-redshift
predictions of galaxy formation models and distinguish between the
traditional ``monolithic collapse'' of proto-galactic gas clouds
(Eggen, Lynden-Bell \& Sandage 1962; Jimenez et al. 1999) and the
hierarchical models of Cold Dark Matter cosmologies (Kauffmann 1996;
Baugh et al. 1996; Cole et al. 2000). Moreover, they allow a powerful
insight in the mechanisms generating the global relations observed
with the structure parameters of early-type galaxies (the
color-magnitude relation, Bower et al. 1992; the Mg-$\sigma$ relation,
Bender, Burstein \& Faber 1993; Colless et al. 1999; the Fundamental
Plane, Djorgovski \& Davis, 1987, Dressler et al. 1987) and regulating
their scatter. In addition, they are the local counterpart of the
studies of the evolution of early-type galaxies with redshift in
clusters (van Dokkum \& Franx 1996, Bender et al. 1998, van Dokkum \&
Standford 2003) and in the field (Treu et al. 1999; van Dokkum et
al. 2001). Analyzing a large number of objects, they are also able to
detect the possibly subtle effects of the environment.

The present paper cannot compete in terms of galaxy numbers with the
literature discussed above. Rather, it relies on the high quality of
the data and stresses two aspects neglected in the past: First a
rigorous quantitative analysis of the \aFe\ role in determining the
ages and metallicities of early-type galaxies, and second the
discussion of the radial variations of the stellar populations of the
galaxies. The former is based on newly developed models of the Lick
indices (TMB) that incorporate the response functions of Tripicco \&
Bell (1995), allowing a proper treatment of the element
abundances. The latter provides an additional tool to test the
predictions of galaxy formation models.

``Classical'' monolithic collapse models with Salpeter IMF (Carlberg
1984) generate steep metallicity gradients (the centers are more metal
rich than the outer parts), shallow positive age gradients (the
centers are slightly younger than the outer parts) and positive
overabundance gradients (the centers are less \aFe\ enhanced, Thomas,
Greggio and Bender 1999). The merger trees typical of hierarchical
models of galaxy formation (Lacey \& Cole 1993) are expected to dilute
gradients originally present in the merging galaxies and therefore
give end-products with milder gradients (White 1980), although
detailed quantitative predictions are still lacking. There is
consensus that the radial changes in colors and line indices observed
in local ellipticals reflect variation in metallicities rather than
age (Faber 1977; Davies et al. 1993; Kobayashi \& Arimoto 1999), a
conclusion confirmed by the study of the evolution of the color
gradients with redshift (Saglia et al. 2000; Tamura et
al. 2000). However, nothing is known about possible radial gradients
of the \aFe\ abundance ratio detected at the centers of early-type
galaxies. If no \aFe\ gradients are present, then the bulk of the
stars of elliptical galaxies are \aFe\ overabundant and the constraint
of short ($\le 1$ Gyr) formation time-scales (e.g., Matteucci 1994;
Thomas, Greggio \& Bender 1999) applies to the stellar populations of
the galaxies as a whole, and not only to the galaxy centers. Our
dataset clarifies this issue.

\medskip
The paper is organized as follows. Section \ref{data} describes the
dataset. Section \ref{indices} discusses the central values and 
the radial profiles of the
line indices. Section \ref{agemetover} derives constraints on the age,
metallicities and the $[\aFe]$ distributions inside the galaxies
(central values and gradients).  Conclusions are drawn in Section
\ref{conclusion}.

\section{The data sample}
\label{data}

The basis of this investigation are the radial line index profiles of
\Mgb, \Fe, and \Hb\ as well as the profile of the velocity dispersion
$\sigma$ derived for 35 early-type galaxies in Mehlert et al. (2000,
Paper~I).  These parameters were obtained from high S/N and spatially
resolved longslit spectra obtained along the major axis of the
galaxies. Details on the galaxy sample, the observations, data
reduction and parameter derivation are described in Paper~I.  With
this dataset we can not only investigate the stellar population of the
galaxies based on their central line indices, but also derive the
gradients of the line indices and the stellar population parameters
like age, metallicity and the $[\aFe]$ ratios.  Since the sample
analyzed here comprises E and S0 galaxies and spans a range of 3 dex
in 
%%%new
%{\bf 
local mass densities (as determined by X-ray observations, Briel, Henry
  and B\"ohringer 1992), 
%}
%%%
the influence of galaxy type and environmental density on
the stellar population of the Coma cluster early type galaxies can be
explored.  In Paper~I we already showed that our derived line index
and $\sigma$ values agree quite well with literature data. This is
also supported by a detailed comparison to a different galaxy sample
in Moore et al. (2002).

%**********************************************************************
\begin{figure}[ht]
% FIGURE 1
%includegraphics*[width=\linewidth]{mgbfehbe_sig.eps}
\includegraphics*[width=.99\linewidth]{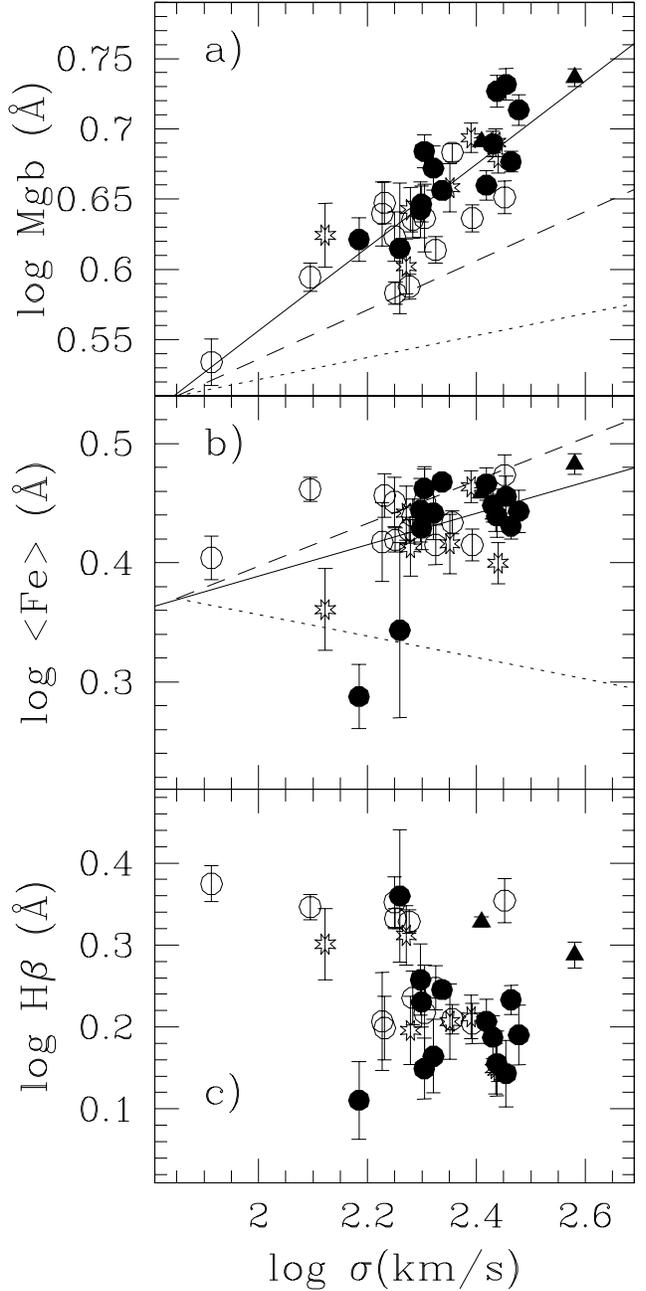}
\caption{The central line indices \Mgb\ (panel a), \Fe\ (panel b),
and \Hb\ (panel c) versus the velocity dispersion $\sigma$ for the 35
early type galaxies of the Coma cluster sample.  Filled triangles
indicate the 2 cDs, filled circles Es, open circles S0s and open stars
E/S0s.  The solid lines are linear least square fits (see text).
%%%new
%{\bf 
The long and short dashed lines show the contributions to the
  Mg$-\sigma$ and Fe$-\sigma$ relations due to metallicity and
  \aFe\ ratio variations, respectively (see Sect. \ref{secveldisp}).  
%}
%%%
}
\label{fig:mgbfehbe_sig}
\end{figure}

\section{Line indices}
\label{indices}

\subsection{Central values}
\label{centralindices}
Table~\ref{tab:indices_zero} lists the central values of the indices
\Hb, \Mgb, Fe5270, Fe5335 and of the velocity dispersion $\sigma$ with
the respective errors.  The values have been derived from the major
axis profiles shown in Paper I averaging data points inside $0.1 a_e$,
where $a_e = r_e / \sqrt{1-\epsilon}$ is the distance along the major
axis, that corresponds to the effective radius $r_e$ of the galaxy
with ellipticity $\epsilon$.  The central values were calculated by
weighting proportional to the signal-to-noise ratio S/N:
\begin{equation}
\langle {\rm Index}\rangle = \frac{\sum_i {\rm Index}_i \times
(S/N)_i} {\sum_i (S/N)_i} \quad ,
\label{eqindex}
\end{equation}
where $i$ indicates the radial points available. This way of weighting
our radial values turned out to be a good compromise to optimize the
quality of the mean values without overemphazising the central
value.

In the following we indicate the average Iron index with $\Fe=({\rm
Fe5270}+{\rm Fe5335})/2$ (Gorgas et al. 1990) and the combined
Magnesium-Iron index with
\begin{equation}
\MgFe=\sqrt{\Mgb\,(0.72\times\mathrm{Fe5270} +
0.28\times\mathrm{Fe5335})}
\end{equation}
\label{eqmgfep}
newly defined by TMB. This index turns out to be completely
independent of the \aFe\ and hence serves best as a metallicity
tracer. The usual index $[{\rm MgFe}]=\sqrt{\Mgb\,\Fe}$ earlier
defined by Gonz\'alez (1993) still depends a little on \aFe\ (see also
Sect.~\ref{model}).  We also consider the $\Mgb^{\prime}$ index (see
Colless et al.\ 1999):
\begin{equation} 
\Mgb^{\prime} = -2.5 \times \log \left(1 - \frac{\Mgb}{\Delta
\lambda}\right),
\label{eqmgbp}
\end{equation} 
with  $\Delta \, \lambda = 32.25$ \AA. 

%**********************************************************************
%**********************************************************************
\begin{figure*} 
\begin{center} 
% FIGURE 2
\includegraphics{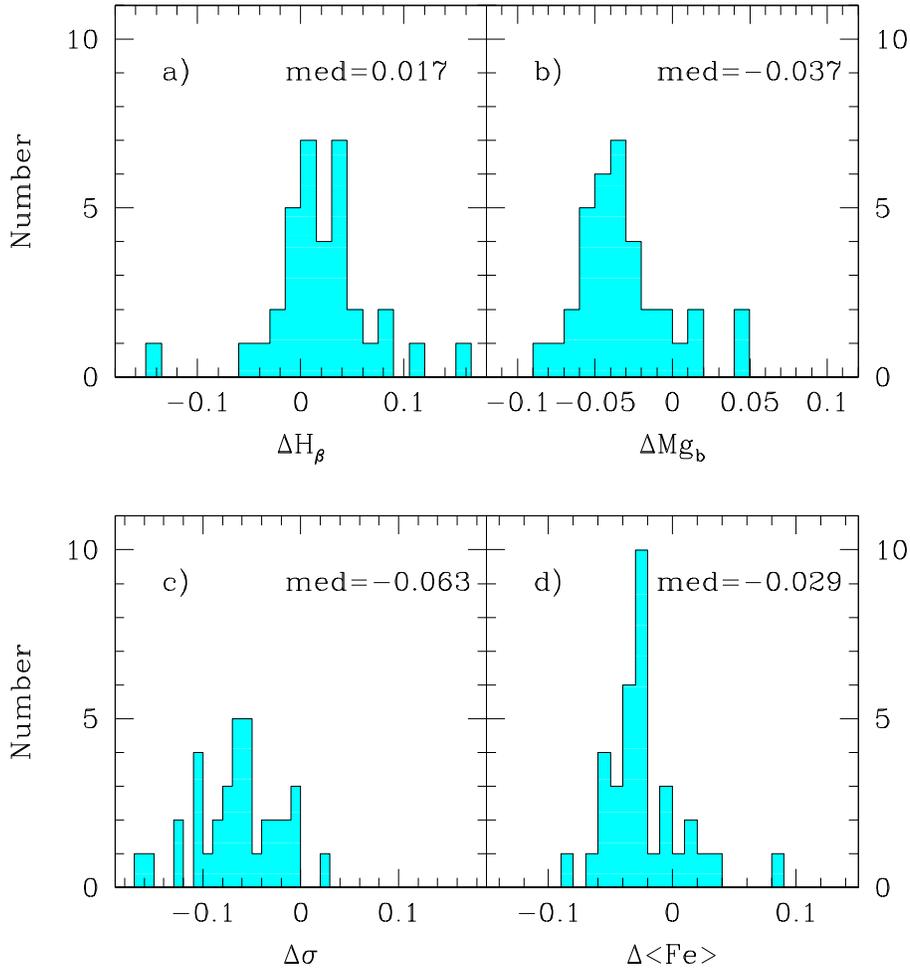} 
\caption{Distributions of the fitted logarithmic gradients of the
indices, \Hb\ (a), \Mgb\ (b), \Fe\ (c), and the velocity dispersion
$\sigma$ (d). The median of the gradients is indicated in the
individual panels.
\label{fig:indgradhist} 
}
\end{center} 
\end{figure*} 
%**********************************************************************

In Fig.~\ref{fig:mgbfehbe_sig} we plot the central index values as
a function of velocity dispersion. Our data show a tight Mg-$\sigma$
correlation (panel a) in good agreement with the literature (e.g.,
Bender et al.\ 1993; Colless et al.\ 1999).  Correlations of velocity
dispersion with the other indices \Fe\ and \Hb, instead, are much less
well-defined (panels b and c) in agreement with results from, e.g.,
J{\o}rgensen et al.\ (1999). Still, \Fe\ seems to be weakly correlated
with $\sigma$ (see also Kuntschner 2000). Linear least square fits
(see the solid lines in Fig.~\ref{fig:mgbfehbe_sig}) yield the
following relations:
\begin{eqnarray}
\Mgb^{\prime} &=& -0.11 (\pm 0.03) + 0.12 (\pm 0.01) \times \log
\sigma\nonumber\\ 
\log\Mgb &=& -0.04 (\pm 0.08) + 0.30 (\pm 0.03) \times \log
\sigma\nonumber\\ 
\log\Fe &=& 0.13 (\pm 0.11) + 0.13 (\pm 0.05) \times
\log \sigma\nonumber\\ 
%\log\Hb &=& 0.78 (\pm 0.22) - 0.23 (\pm 0.09) \times \log\sigma
\label{eqmgbsig}
\end{eqnarray}

Although there is no striking difference between elliptical and
lenticular galaxies in Fig.~\ref{fig:mgbfehbe_sig} (see also Mehlert
1998), we note that the S0s in our sample have on average lower
central velocity dispersions, and tend to have slightly lower metallic
indices at a given $\sigma$. The most significant feature is the
separation of the lenticulars in two distinct subclasses, one being
almost indistinguishable from the ellipticals, the other having
extraordinarily high \Hb\ values. We will come back to this point in
Sect.~\ref{agemetover}, where we discuss the derived ages and
element abundances.

Note that also the two cD galaxies (triangles) have Balmer lines that
are unusually strong given the relatively high galaxy's velocity
dispersion.  For a detailed discussion of the properties of the
stellar populations (age, metallicity and \aFe\ ratios) derived from
these indices see Sect.~\ref{agemetovercentral}.

Finally it should be emphasized that we do not find any dependence of
the derived indices or the galaxies velocity dispersion on the density
profile of the Coma cluster.

\subsection{Gradients}
\label{indexgradients}

We compute logarithmic index and velocity dispersion gradients
performing a linear $\chi^2$ fit to the data points within $a<a_e$,
where $a$ is the distance from the center along the major
axis.
\begin{equation}
\mathrm{\triangle}({\rm Index}) = \frac{\Delta \log({\rm
Index})}{\Delta \log (a/a_e)}
\label{eqgrad}
\end{equation}
The fitted gradients and the logarithmic values of the indices and
$\sigma$ at $1 a_e$ are listed in Tables~\ref{tab:indices_gradients}
and \ref{tab:indices_re_fit}, respectively. The number distribution of
the gradients and their median values are shown in
Fig.~\ref{fig:indgradhist}.  The velocity dispersion $\sigma$ and the
metallic line indices \Mgb\ and \Fe\ show on average negative
gradients. The median of the Balmer line gradient, instead, is
positive.  The $\sigma$-gradients measured here (with a median values
of -0.063) agree quite well with
those found by Davies et al. (1987) and are slightly steeper than the 
modeling of J\o rgensen et
al. (1995), who use -0.04.  
\Mgb\ gradients are not available in the literature, the
Mg$_2$ gradient measured by Davies et al. (1993; $\triangle$Mg$_2 =
-0.059 \pm 0.022$), however, tends to be slightly steeper than our
values ($-0.037$ for $\log\Mgb$, see Fig.~\ref{fig:indgradhist}).  The
median \Fe\ gradients in Fig.~\ref{fig:indgradhist} are also somewhat
shallower than measurements from Davies et al.\ (1993) and Gorgas et
al.\ (1990), but are consistent within the errors. Reliable \Hb\
gradients for meaningful comparisons are not available in the
literature.

Interestingly, we do not find any statistically significant
correlation between index gradients and their central values or central
velocity dispersion. There is only a very weak hint, that galaxies
with stronger central metallic lines have steeper index gradients.
Similar results for the Mg$_2$ index have been found for 114 and 42
elliptical galaxies in the field by Gonz\'alez \&\ Gorgas (1995) and
Carollo, Danziger, \&\ Buson (1993), respectively. 

%**********************************************************************
\begin{figure} 
% FIGURE 3
\includegraphics[width=8.7cm]{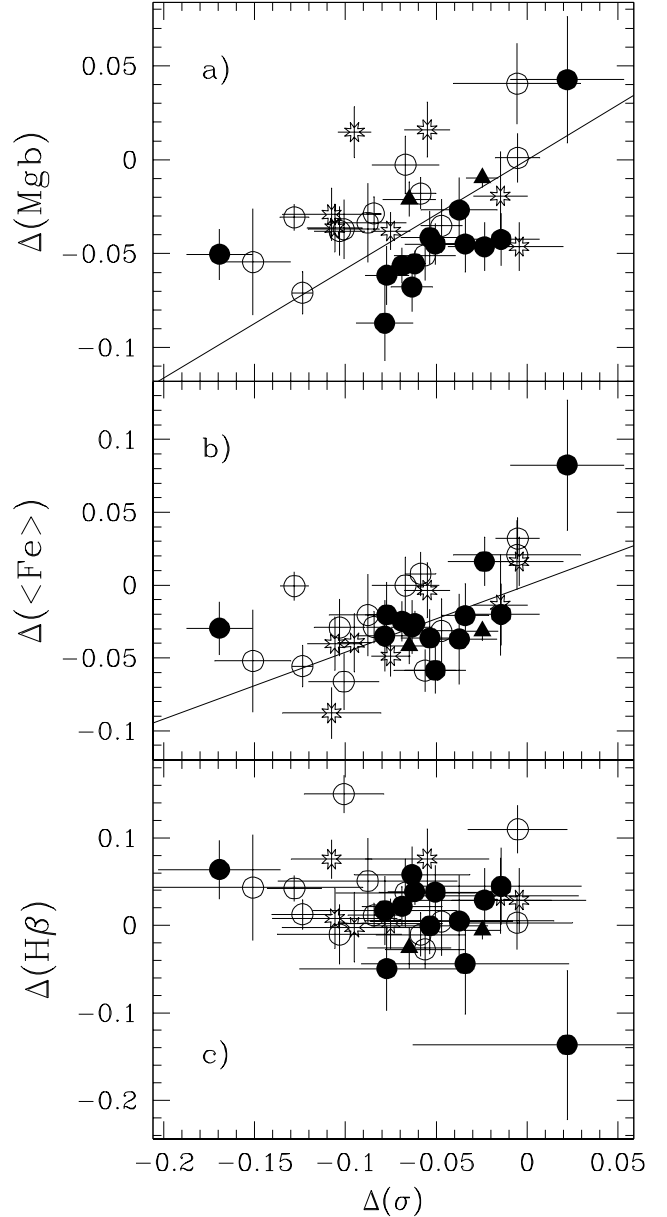} 
\caption{The fitted logarithmic gradient of the Mg$_b$ {\bf (a)}, $<$Fe$>$
{\bf (b)} 
%%%new
%{\bf 
and H$\beta$ {\bf (c)} 
%}
%%%
index versus the
fitted logarithmic gradient of the velocity dispersion $\sigma$. 
Symbols as in Fig.~\ref{fig:mgbfehbe_sig}. The lines show the
relations  $\langle\triangle \Mgb\rangle \approx 0.58 \times
\langle\triangle\sigma\rangle$ and  
$\langle\triangle \langle Fe\rangle\rangle \approx 0.46 \times
\langle\triangle\sigma\rangle$ (see Fig. \ref{fig:indgradhist}). 
\label{fig:gradmgbfe_gradsig} 
} 
\end{figure} 
%**********************************************************************
We do find a correlation, however, between the metallic index
gradients and the gradients in velocity dispersion. In
Fig.~\ref{fig:gradmgbfe_gradsig} it is shown that both $\triangle\Mgb$
and $\triangle\Fe$ become steeper with a steepening of
$\triangle\sigma$ at the significance level of $\approx 3\sigma$ and
$5 \sigma$, respectively.  Hence we basically confirm the validity of
the assumption, made by J{\o}rgensen et al.\ (1995) when modeling
aperture corrections, that the \Mgb\ and \Fe\ gradients correlate with
the gradient of $\sigma$. However, this does not hold for \Hb\
(see~\ref{fig:gradmgbfe_gradsig}c), where the median gradient is
slightly positive. Moreover, our results indicate an aperture
correction slightly steeper (but consistent within the errors) than
the one proposed by J{\o}rgensen et al.\ (1995) for the $\sigma$
measurements ($\sigma_{ap}/\sigma_{norm}=(r_{ap}/r_{norm})^{-0.06}$
instead of ($\sigma_{ap}/\sigma_{norm}=(r_{ap}/r_{norm})^{-0.04}$), a
shallower one for the \Fe\ values
($\Fe_{ap}/\Fe_{norm}=(r_{ap}/r_{norm})^{-0.03}$), while on the spot
for the \Mgb\ data
($\Mgb_{ap}/\Mgb_{norm}=(r_{ap}/r_{norm})^{-0.04}$).  Anyway, in a
recent paper Moore et al.\ (2002) show that the aperture correction is
usually negligible since it is typically smaller or at most of the
order of the individual index errors. It is also interesting to note
that the slope of the {\it local} $\triangle\Mgb-$$\triangle\sigma$
correlation is not very different from the one of the {\it global}
correlation of Eq. \ref{eqmgbsig}.
   
Finally we note that, like for the central index values, no correlation
between the index gradients and the density profile of the Coma
cluster is present in our galaxy sample.

\section{\boldmath Ages, metallicities and $[\aFe]$ ratios}
\label{agemetover}

\subsection{The model}
\label{model}

In the following we use the stellar population models of TMB in order
to derive the stellar
population parameters age, metallicity, and \aFe\ ratio from the line
indices \Hb, \Mgb, and \Fe .

%**********************************************************************
\begin{figure*} 
% FIGURE 4
\includegraphics[width=0.49\linewidth]{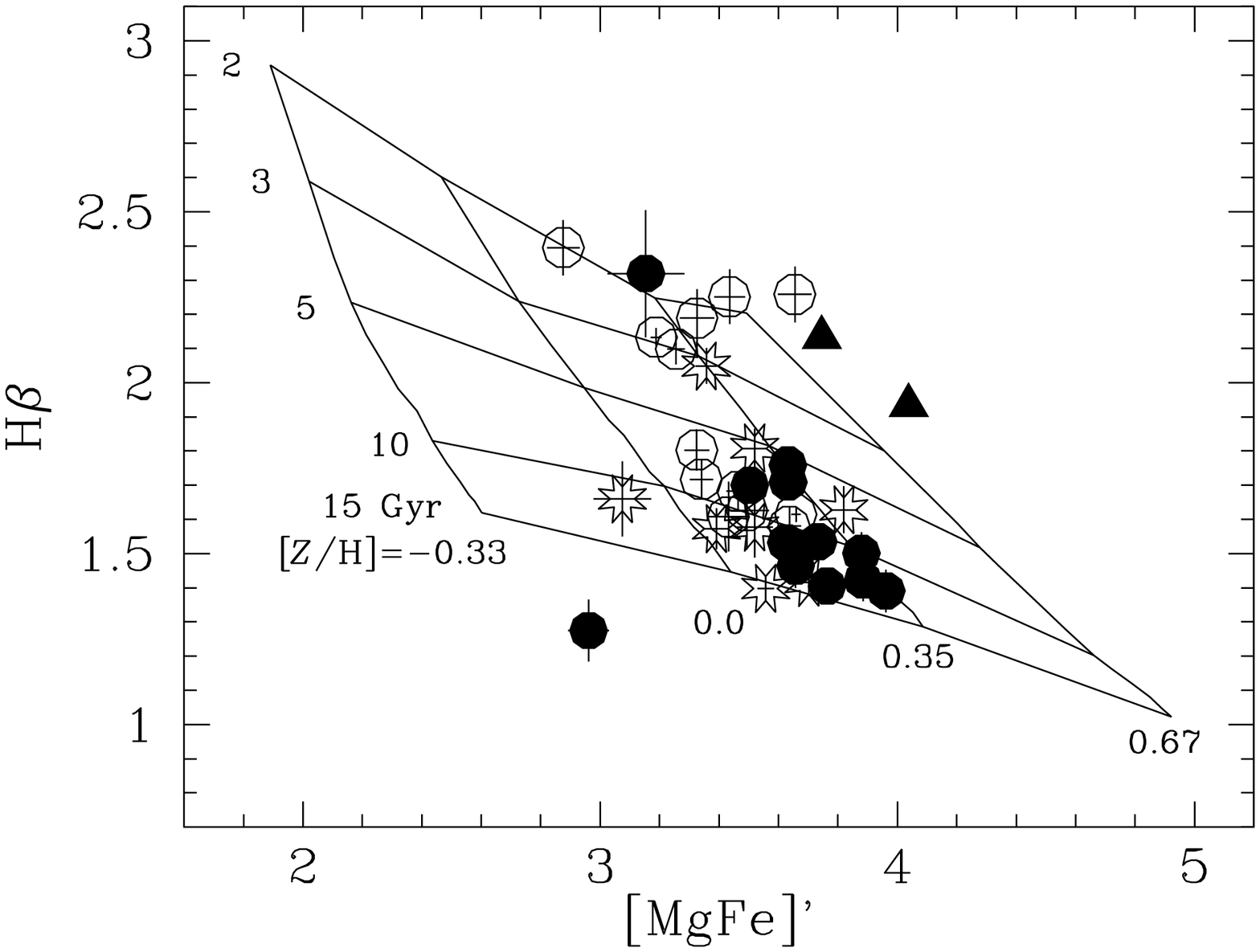} 
\includegraphics[width=0.49\linewidth]{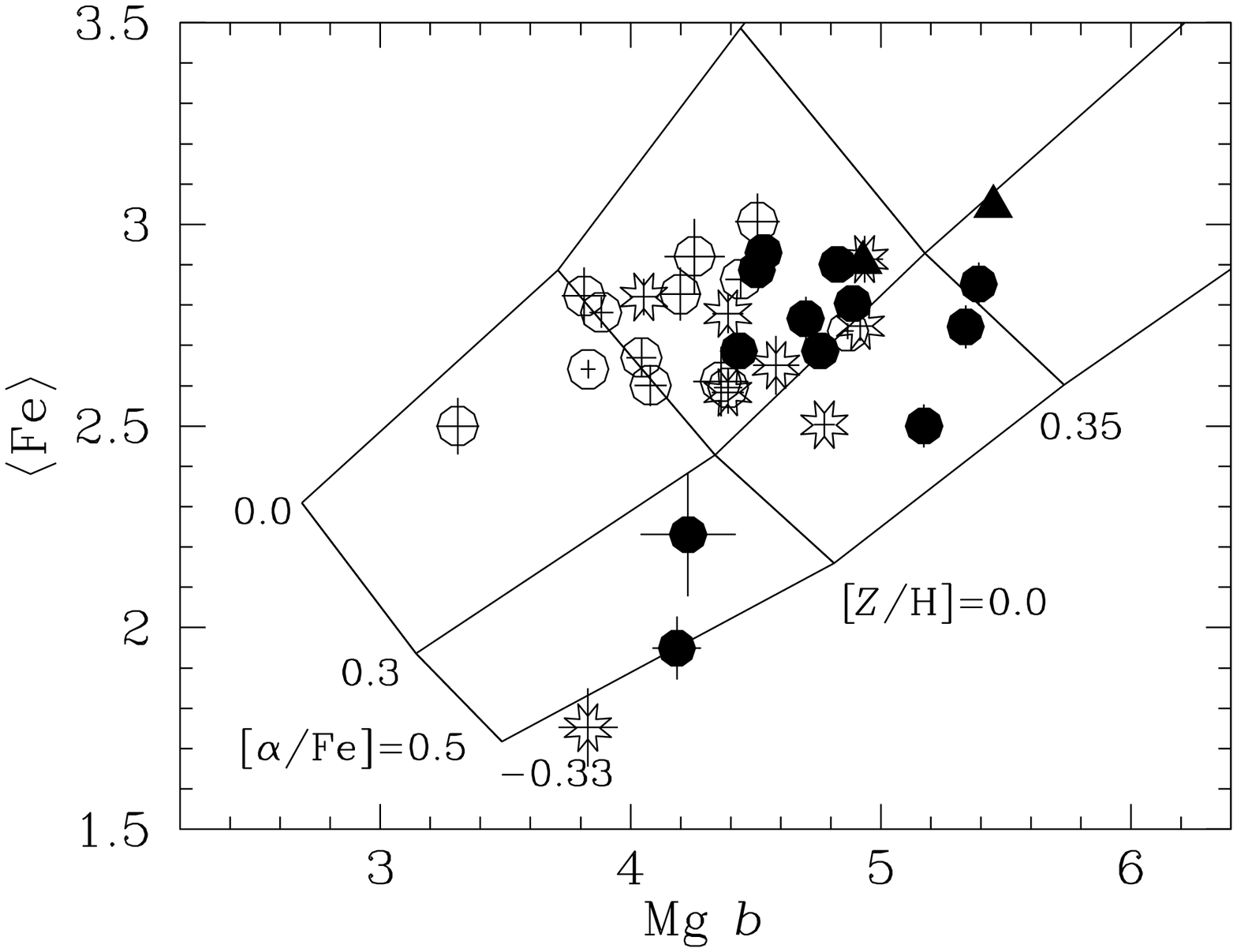} 
\caption{The distribution of the central values of \Hb\ and \MgFe\
indices (left panel) and \Fe\ and \Mgb\ indices (right panel) averaged
over 0.1 $a_e$ of the 35 early type galaxies of the Coma cluster
sample.  Symbols as in Fig.~\ref{fig:mgbfehbe_sig}. The lines indicate
models (TMB) of constant ages and metallicities (see labels). Models in
the right panel are for fixed age (12~Gyr) and various \aFe\ ratios
(see labels).
\label{fig:mgfehb} 
} 
\end{figure*} 
%**********************************************************************
The TMB models take into account the effects on
the Lick indices from element abundance ratio changes, hence give Lick
indices of simple stellar populations not only as a function of age and
metallicity but also as a function of the \aFe\ ratio. They are based
on the evolutionary population synthesis code of Maraston (1998). The
impact from element ratio changes is computed with the help of the
Tripicco \& Bell (1995) response functions, using an extension of the
method introduced by Trager et al.\ (2000a).  Because of the inclusion
of element ratio effects, the models allow for the clear distinction
between total metallicity [\ZH] and the $\alpha$ to iron-peak elements
ratio [\aFe]. The latter can be best derived from the classical Lick
indices \Mgb\ and Fe5270 and/or Fe5335 (Maraston et al.\ 2003, TMB).
The models of TMB cover the ages 1 to 15 Gyr, metallicities $-2.25\leq
[\ZH]\leq 0.67$, and the abundance ratios $-0.3\leq [\aFe]\leq 0.5$.

%%%new
%{\bf 
The Balmer line \Hb\ is widely used as an age indicator to break
the age-metallicity degeneracy (Gonz\'alez 1993; Worthey 1994),
because it is very sensitive to warm turnoff stars.  We note the
caveat, however, that the possible presence of unusually warm
horizontal branch stars in the stellar population can hamper the
usefulness of Balmer lines as age indicators and seriously affect the
age diagnostics (Maraston \& Thomas 2000).  Such blue horizontal
branch stars may originate from either a metal-poor subpopulation or
metal-rich populations with blue horizontal branch morphologies due to
enhanced mass loss along the red giant branch evolutionary phase.  The
TMB models are constructed with the canonical mass loss rates along
the red giant branch (Fusi-Pecci \& Renzini 1976), that have been
calibrated on galactic globular clusters (Maraston \& Thomas 2000;
Maraston et al.\ 2003).  The effect described above, i.e.\ anomalies
in the horizontal branch morphology as a further parameter besides
age, metallicity, and \aFe\ ratio, is not considered in the present
paper.

We derive the three stellar population parameters age, metallicity,
and \aFe\ ratio from the three line indices \Hb, \Mgb, and \Fe\ in a
twofold iterative procedure. First, we arbitrarily fix the \aFe\
ratio, and determine ages and metallicities for the index pairs
(\Hb,\Mgb) and (\Hb,\Fe), by starting with arbitrary age-metallicity
pairs, which we modify iteratively until both index pairs are
reproduced. The two metallicities obtained from \Mgb\ and \Fe,
respectively, are used to adjust the \aFe\ ratio, and to start a new
iteration. These steps are repeated until the age-metallity pairs
derived from (\Hb,\Mgb) and (\Hb,\Fe) at a given \aFe\ ratio are
consistent within 1 per cent accuracy. For ages and metallicities
between the grid points quoted above, we interpolate linearly.
%}
%%%new

\subsection{Central values}
\label{agemetovercentral}

From the central line indices (Table~\ref{tab:indices_zero}) we now
derive the central  ages, metallicities and \aFe\ ratios of the
stellar population for the early type galaxies investigated in this
paper using the TMB stellar population model described in Sect.~\ref{model}. 

\subsubsection{The diagnostic diagrams}
\label{diagrams}

The left diagram in Fig.~\ref{fig:mgfehb} shows the distribution of
the central \Hb\ and \MgFe\ indices plotted with the stellar
population models of TMB.  In this parameter space, mean age and total
metallicity can be best read-off by eye, as both indices are
practically insensitive to \aFe\ ratio variations as shown in TMB.  In
the right-hand panel of Fig.~\ref{fig:mgfehb} we show models and data
in the \Mgb-\Fe\ plane, which provides a reasonably good first
approximation of \aFe\ ratios, although age effects cannot be
neglected for the final derivation of \aFe\ ratios. The models in the
right panel are plotted for fixed age (12~Gyr).  The precise central
ages, metallicities and \aFe\ ratios derived for our sample
are listed in Table~\ref{tab:ages_zero}.

Most objects have super-solar total metallicities and \aFe\ element
ratios, hence they are \aFe\ enhanced. The sample spans a relatively large
range in average ages from 2 to 15~Gyr in agreement with a recent study
of Coma cluster galaxies by Trager et al.\ (2003). The
bimodal distribution in the \MgFe-\Hb\ parameter space is particularly
interesting. We notice one
'clump' at old ages scattering about an average age of 10~Gyr with a
rather narrow distribution in metallicities ($0\leq [\ZH]\leq 0.35$),
and a second one at roughly 2~Gyr spanning a larger range in
metallicities ($0\leq [\ZH]\leq 0.8$).  There is a very clear
separation between the two. While the 'old clump' contains all types
of objects, i.e.\ ellipticals (E), lenticulars (S0) and transition
types (E/S0), the 'young clump' is by far dominated by S0 galaxies
plus the two cDs. This result was already indicated by the
distribution of the \Hb\ indices discussed in
Sect.~\ref{centralindices},
where a subclass of lenticulars with high \Hb\  was pointed out.

\begin{figure*} 
\begin{center} 
% FIGURE 7
\includegraphics{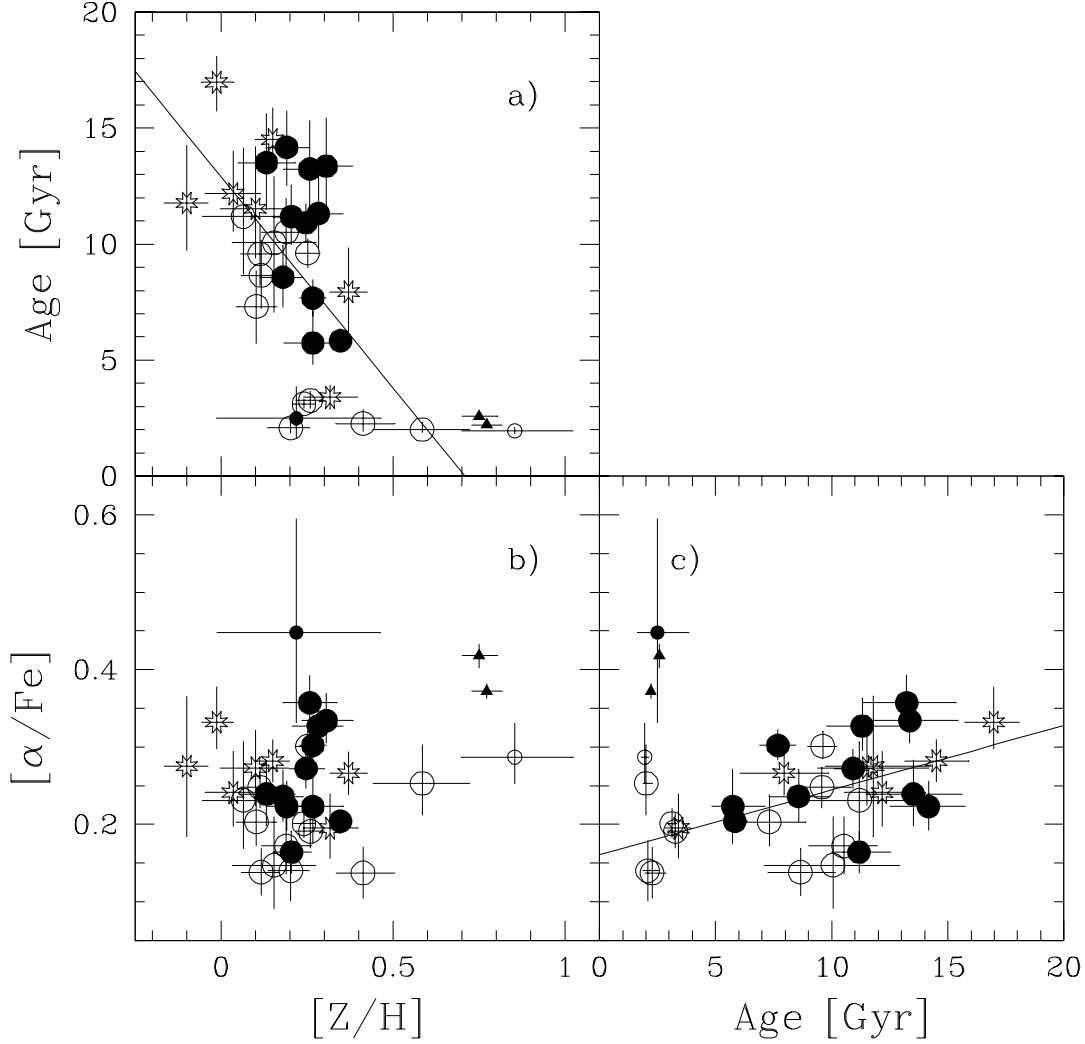} 
\caption{Central ages, metallicities ($[Z/$H$]$) and the $[\aFe]$
ratios listed in Tab.~\ref{tab:ages_zero} plotted versus each
other. Symbols as in Fig.~\ref{fig:mgbfehbe_sig}. The linear
least square fits for panel a) and c) give following relations a)
$\mathrm{age} = 12.9 (\pm 1.2) -18.2 (\pm 4.7) \times
[Z/\mathrm{H}]$ and b) $ [\aFe] = 0.16 (\pm 0.02) + 0.008 (\pm 0.002)
\times \mathrm{age}$ and are indicated by the straight lines. Object 
GMP 3329, 2921, 3561 and 3201 (small symbols) were not included in the 
fit 
%%%new
%{\bf 
(see Sect.~\ref{diagrams}).
%}
%%%new
\label{fig:age_met_a_fe} 
} 
\end{center} 
\end{figure*}

\begin{table*}
\caption
{Type dependent mean values and 1 $\sigma$ rms scatter of the central
ages, metallicities ($[Z/$H$]$) and $[\aFe]$ ratios.
\label{tab:meanages_type}}
\centerline{
\begin{tabular}{crcccccc}
\hline\hline
Type& N & $<$age$>$&rms(age)&$<[Z/$H$]>$&rms($[Z/$H$]$)&$<[\aFe]>$&rms($[\aFe]$)\\
& & $[$Gyr$]$&$[$Gyr$]$& & & & \\
\hline
    E &  11 &   10.5 &   3.1 &  0.24 &  0.06 &  0.26 &  0.06 \\
 E/S0 &    7  &    11.2   & 4.4  &  0.12 & 0.17 & 0.27 & 0.04\\
  S0$_1$ &   7 &   9.6 &   1.3 &  0.14 &  0.06 &  0.21 &  0.06  \\
  S0$_2$ &   5  &   2.6  &  0.6  & 0.34 &  0.16 &  0.18 &0.05 \\
\hline                                     
\end{tabular}
}
\end{table*}
The average stellar parameters of these subclasses are summarized in
Table~\ref{tab:meanages_type}. The averages exclude, however, the
galaxy GMP 3958 (E) because of its extremely low Balmer index
($\Hb=1.29$, see Table~\ref{tab:indices_zero}), which implies an
extrapolated unreasonable age of 24 Gyr (see Fig.~\ref{fig:mgfehb}).
Also the three objects GMP3329 (cD), GMP2921 (cD), and GMP3561 (S0)
are not included in these average values, because they lie outside the
model grid, so that their derived stellar parameters are strongly
affected by uncertainties caused by extrapolation. 
%%%new
%{\bf 
Finally, we excluded galaxy GMP3201 from the further discussion
(averages in Table~\ref{tab:meanages_type} and linear fits in
Figs.~\ref{fig:sp_sig} and~\ref{fig:age_met_a_fe}), since the errors of
the measured indices and hence the corresponding stellar parameters are
systematically larger than those of all other objects.
%}
%%%new

Table~\ref{tab:meanages_type} confirms the S0 dichotomy found by
Poggianti et al. (2001a), who investigated a sample of 52 Coma early
type galaxies. They interpret the young class (S0$_2$) as being the
descendants of typical star-forming spirals whose star formation has
been stopped due to the dense cluster environment, while the
evolutionary history of the S0$_1$ class and Es should be identical.
%%%new
%{\bf 
The low average ages of the young S0 subclass (S0$_2$) measured
would then be the direct consequence of recent star formation that had
ceased just recently due to the infall in the dense cluster
environment. The resulting extended star formation histories would
imply, however, a significant enrichment of iron from Type Ia
supernovae, so that low \aFe\ ratios should be observed in contrast to
the results found here (see Table~\ref{tab:meanages_type}). The
relatively high \aFe\ ratios disfavor the occurence of recent star
formation. They would be 
%%%gary
%{\bf 
more
%}
 compatible with the alternative
interpretation that the high Balmer line indices measured in the
objects of class S0$_2$ are actually caused by the presence of
unusually blue horizontal branches in these objects rather than by
young stellar populations (see Sect.~4.1).
%}
%%%new

\subsubsection{Correlations with velocity dispersion}
%**********************************************************************
\label{secveldisp}
\begin{figure}
\includegraphics*[width=0.8\linewidth]{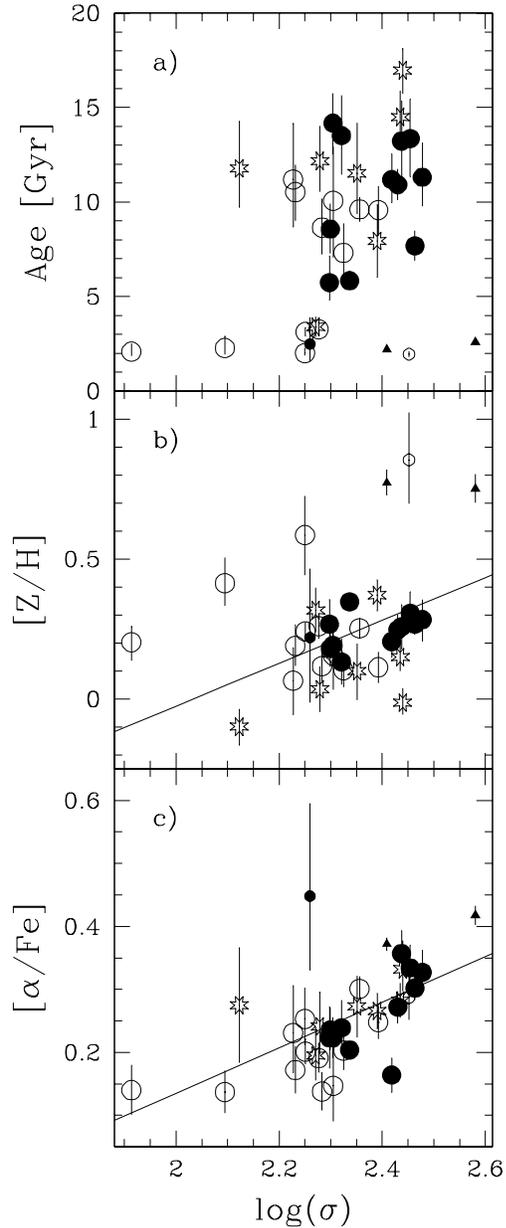} 
\caption{Central ages, metallicities ($[Z/$H$]$) and the $[\aFe]$
ratios listed in Tab.~\ref{tab:ages_zero} plotted versus the central
velocity dispersion $\sigma$.  Symbols as in Fig.~\ref{fig:mgbfehbe_sig}.
%%%new 
%%%gary
%{\bf 
Straight lines indicate the linear least square fits for panels
  b) (excluding the
low-age, high metallicity objects) and c) (excluding the Objects GMP
3329, 2921, 3561 and 3201, small symbols, see Sect.~\ref{diagrams}),
giving the following relations: $[\mathrm{Z/H}] = -1.56 (\pm 0.27) +
0.77 (\pm 0.10) \times \log(\sigma)$, $[\mathrm{\alpha/Fe}] = -0.59
(\pm 0.21) + 0.36 (\pm 0.09) \times \log(\sigma)$.
%}
%%%
}
\label{fig:sp_sig} 
 
\end{figure} 
%**********************************************************************
Fig.~\ref{fig:sp_sig} shows the derived stellar parameters as
a function of velocity dispersion $\sigma$. The parameter which
correlates best with $\sigma$ is the element ratio \aFe\ (panel
c). Excluding the class of young S0s and cDs, also total metallicity
correlates reasonably well with velocity dispersion (panel b). Age,
instead, does not seem to be a major parameter among early-type
galaxies (see also Trager et al.\ 2000b; Kuntschner et al.\ 2002;
Trager et al.\ 2003). Again, the separation in two classes, the
'young' one being dominated by S0 galaxies, is very striking (panel
a). 

It should be emphasized, however, that the age derivation is most
affected by observational errors.  A proper analysis to decide about
possible trends of age with galaxy mass therefore requires a larger
galaxy sample and the consideration of errors, e.g., via Monte Carlo
simulations. Such an analysis is carried out in Thomas, Maraston, \&
Bender (2003b) for a sample of 126 field and cluster galaxies
(including the present sample). Interestingly, the dichotomy mentioned
above is confirmed for this larger sample (including galaxies in low
density environments), and the data of the 'old clump' are best
consistent with age slightly increasing with galaxy mass. Still, the
above conclusion that age is only a secondary parameter remains. The
existence of a [\ZH]-$\sigma$ correlation, instead, is further
reinforced by the study of Thomas et al.\ (2003b).

%%%new
%{\bf 
Fig.~\ref{fig:sp_sig} shows that the elliptical galaxies and the
'old' lenticular galaxies (class S0$_1$ in
Table~\ref{tab:meanages_type}) follow the same correlations of
metallicity and \aFe\ ratio with velocity dispersion. The total
metallicities of the young lenticulars (class S0$_2$ in
Table~\ref{tab:meanages_type}), instead, are significantly ($\sim 0.4$
dex) higher than they are expected to be for their $\sigma$, while
their \aFe\ ratios are consistent with their velocity dispersions. 
% }
%%%new

%%%new
%{\bf 
Using the mean correlations shown in Fig.~\ref{fig:sp_sig}
(assuming a constant age of 9~Gyr) we can now reproduce the \Mgb\ and
\Fe\ indices with the TMB models, and in this way estimate the
relative importance of metallicity and \aFe\ variations in generating
the global Mg$-\sigma$ and Fe$-\sigma$ relations. The resulting
index-$\sigma$ relations for models in which only metallicity or only
\aFe\ increases with $\sigma$ are shown in Fig.~\ref{fig:mgbfehbe_sig}
by the long and short-dashed lines, respectively. We deduce the
contributions from metallicity and \aFe\ variations to the Mg$-\sigma$
relation to be approximately 70 and 30 per cent, respectively. Hence,
we confirm that metallicity is the main driver of the relation, as
suggested in the past. We show, however, that the \aFe\ ratio does
play a non-negligible role. Fig.~\ref{fig:mgbfehbe_sig} further
illustrates that the shallower Fe$-\sigma$ is the result of the
combination of the (positive) metallicity dependence with the {\it
negative} \aFe\ (anti-) correlation. 
%}
%%%

%**********************************************************************
%**********************************************************************
\subsubsection{Correlations between the stellar population parameters}

Possible correlations between the three stellar population parameters
are explored in Fig.~\ref{fig:age_met_a_fe}. There is a weak
indication for a possible age-metallicity anti-correlation among the
old subclass (above $\sim 5$~Gyr). It is more likely, however, that an
anti-correlation at such low significance is an artifact caused by
correlated errors in age and metallicity (Trager et al. 2000b;
Kuntschner et al.\ 2002; Thomas et al.\ 2003b). Note also that the
young subclass is not restricted to high metallicities, but exhibits,
instead, a large range in metallicities from $[\ZH]=0.2$ to 0.9 dex.

We detect the trend that galaxies with higher \aFe\ ratios tend to
have older average ages (panel c, see also Thomas et al.\ 2002). This
result is important, as it supports the connection between \aFe\ ratio
and formation timescale. It is important to take into account that
correlated errors actually lead to the opposite trend, namely higher
\aFe\ ratios at younger ages (Thomas et al.\ 2003b).

It does not come as a surprise that, like the central indices, also
the central ages, metallicities and $[\aFe]$ ratios derived for our
galaxy sample do not correlate with the density profile of the Coma
cluster. Note that, except for a very small fraction of young field
elliptical galaxies (see also Kuntschner et al.\ 2001), also Thomas et
al.\ (2002) could not find any significant differences between
early-type galaxies in low and high density environments with respect
to their stellar parameters.  Environmental density seems indeed to be
only of secondary importance for the formation of early-type galaxies.

\subsection{Gradients}
\label{agemetovergradients}

%**********************************************************************
\begin{figure*} 
% FIGURE 9
\includegraphics[width=18cm]{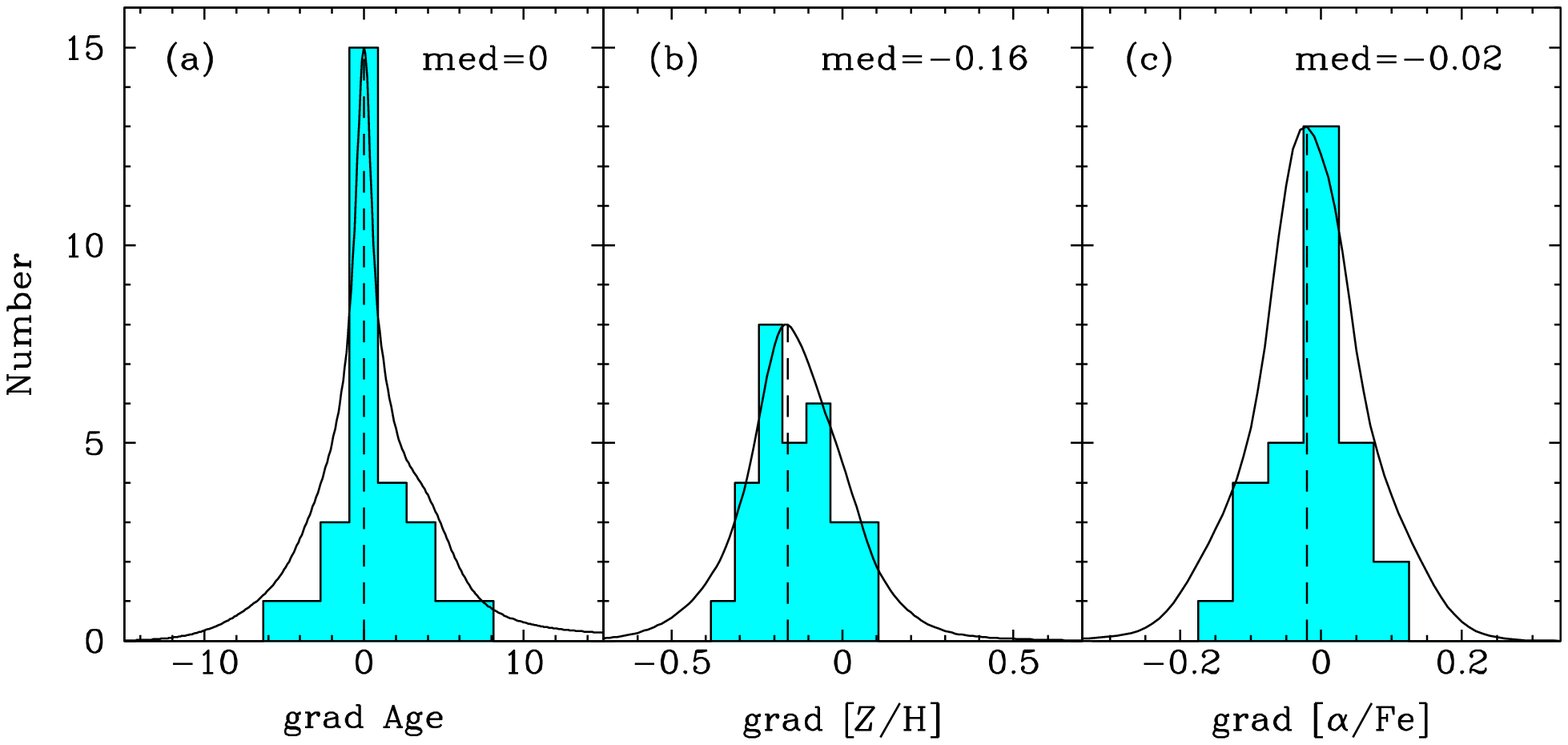} 
\caption{Distributions of the gradients derived for age (Gyr, panel a),
metallicity (panel b), and \aFe\ ratio (panel c). The median values
are given in the panels, respectively. The solid lines are obtained by
summing up the distributions of the gradients folded with their
Gaussian errors.
\label{fig:gradhisto} 
} 
\end{figure*} 
%**********************************************************************
We determine the gradients in ages, metallicities, and \aFe\ ratios
from the fitted index gradients (\Hb, \Mgb, \Fe) and the logarithmic
fit values at $1~a_e$ given in Tables~\ref{tab:indices_gradients}
and~\ref{tab:indices_re_fit} by means of the SSP models of TMB. The
ages, metallicities, and \aFe\ ratios at $0.1~a_e$ and $1~a_e$ are
derived and the gradient was set as the difference between the two.
Hence, we obtain the variation of the stellar population parameters
per decade within $1~a_e$ (see also Equation~\ref{eqgrad}). The
1-$\sigma$ errors were evaluated through Monte Carlo simulations
taking into account the errors in the gradients and the logarithmic
fit values of the three indices. For the two objects GMP3661 and
GMP4679 no reliable gradients could be obtained, because the very high
\Hb\ indices at $1~a_e$ would require extrapolation to ages much
younger than covered by the grid of the present SSP models, which
makes a reliable age determination impossible. Like in case of the
central values, the very low \Hb\ of GMP3958 again made its age
determination unreliable.

The final gradients of the age, metallicity and the $[\aFe]$ ratio
including their errors are listed in Tab.~\ref{tab:ages_gradients}.
Histograms of their number distributions and the median values are
shown in Fig.~\ref{fig:gradhisto}. The solid lines are obtained by
summing up the distributions of the gradients folded with their
Gaussian errors.

\subsubsection{Age and \aFe\ ratio}
%%%new
%{\bf 
The most striking result is that, on average, the early-type
galaxies in the Coma cluster do not exhibit gradients in age or \aFe\
ratio. The number distributions of the age and \aFe\ ratio gradients
(panels (a) and (c) in Fig.~\ref{fig:gradhisto}) show both a very
clear peak at zero. As the distributions in age and \aFe\ ratio are
not well-defined Gaussians, we approximate their rms (Gaussian widths
$\sigma$) by the deviation of 68 per cent of the objects from the
median value of the distribution. The resulting values for the age and
\aFe\ gradients are 2.7~Gyr and 0.05~dex, respectively, and fully
consistent with the respective median errors of the gradients (2.9~Gyr
and 0.05~dex, see Table~\ref{tab:ages_gradients}). In both cases, 91
per cent of the objects fall inside $2\sigma$ of the distributions. To
conclude, the non-detection of a gradient within one effective radius
in both age and \aFe\ for the sample studied here is a statistically
significant result. Devations from the median value can be explained
by observational errors alone. It is worth noting that we do not find
a correlation between grad Age and grad [\aFe], or between grad Z and
grad [\aFe], which further supports the above conclusion. There is a
weak anticorrelation between grad Age and grad Z as a consequence of 
the expected correlated errors. 
%}
%%%new

The lack of gradients in age and \aFe\ ratio is particularly
interesting as both parameters reflect formation timescales. The
presence of \aFe\ enhanced stellar populations is obviously not
restricted to galaxy centers. Early-type galaxies are {\em globally}
\aFe\ enhanced, at least within their effective radii. Hence, the
formation epochs and timescales of the stellar populations do not
significantly change within the galaxy as a function of radius. This
conclusion goes along well with the lack of a significant age
gradient. The stellar populations in a galaxy may form on long or on
short timescales, but they form simultaneously in the entire galaxy
independent of their distance to the galaxy center. Note that also the
lack of a significant evolution with redshift of color gradients in
early-type galaxies strongly disfavors the presence of age gradients
in early-type galaxies (Saglia et al. 2000; Tamura et al. 2000).

This implies that scenarios with strong inside-out or outside-in
formation processes are both disfavored. No significant residual star
formation can have occurred either in the outskirts or in the center
of a galaxy. This stands in conflict with simple monolithic models,
that predict the onset of galactic winds to occur first at large
radii, so that a positive gradient in \aFe\ should be detected
(Martinelli et al. 1998; Thomas et al. 1999).

On the other hand, current models of hierarchical galaxy formation
cannot easily accommodate globally \aFe\ enhanced stellar populations
(Thomas \& Kauffmann 1999).  In general, one expects significant star
formation to rapidly occur in the galaxy center triggered by a merger
event (Schweizer 1990; Barnes 1992; Bender \& Surma 1992). Therefore
an \aFe\ overabundant stellar population should be present there
(i.e. a centrally localized negative gradient, Thomas 1999).
%%%new
%{\bf 
However, note that the spatial resolution of our data (typically
$\sim 2''$, see Mehlert et al.\ 2000) allows us to detect decoupled
cores in only two of our 35 galaxies (see Mehlert et al. 1998) and
that our \aFe\ gradients avoid the central regions. 
% }
%%%

%If the \aFe\ overabundant
%population were to be formed during the merger event, a negative
%gradient in \aFe\ would be expected (Thomas 1999, Thomas et al.\
%1999), in conflict with the observational results found here. If a
%significant amount of star formation is supposed to happen in mergers,
%the gas (or at least the newly formed stellar populations) must
%distribute more uniformly than expected from simulations.

\subsubsection{Metallicity}

\begin{figure*} 
% FIGURE 10
\includegraphics[width=0.49\linewidth]{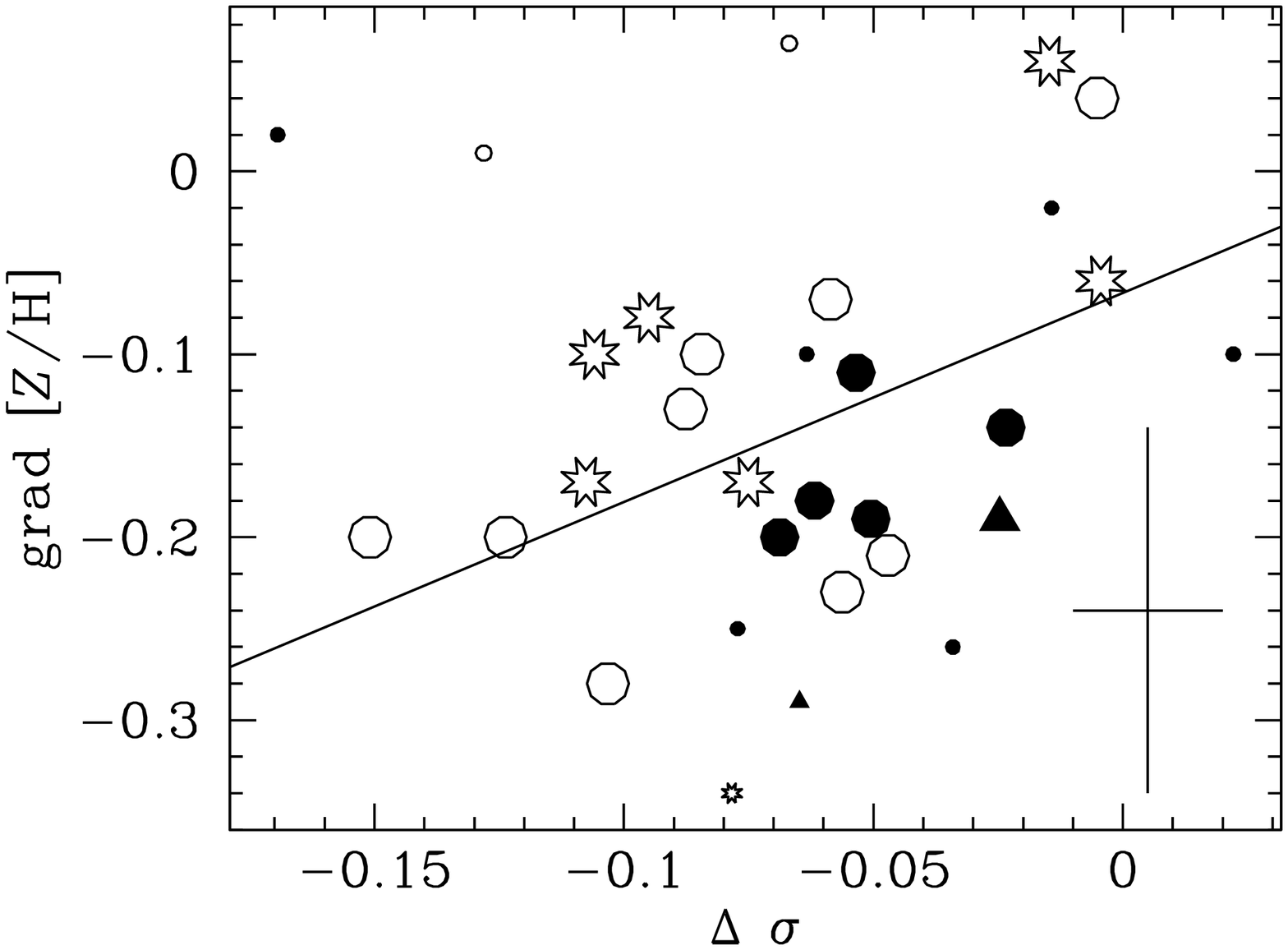} 
\includegraphics[width=0.49\linewidth]{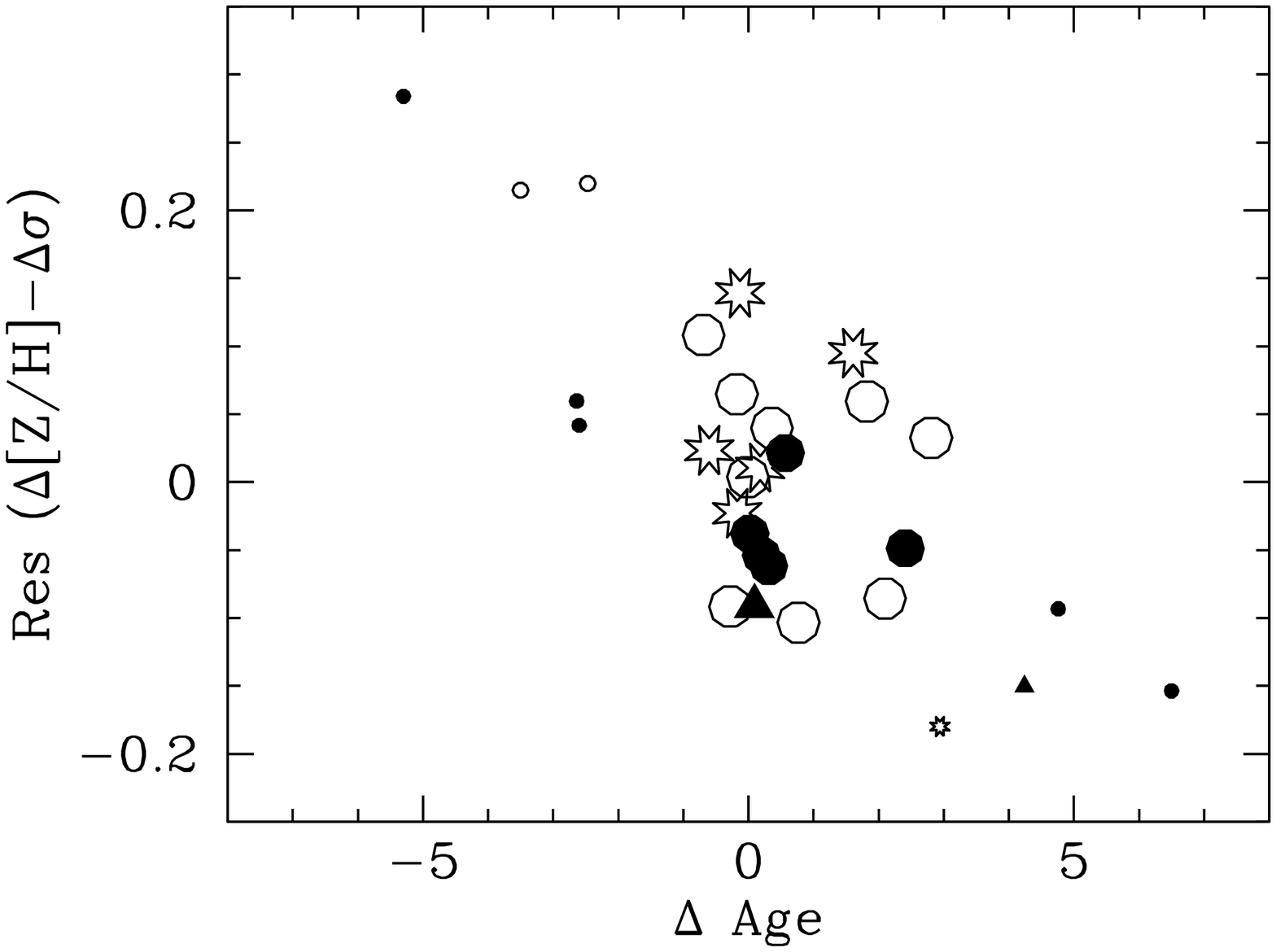} 
\caption{{\em Left panel:} Metallicity gradient as a function of the
logarithmic gradient in velocity dispersion. Symbols as in
Fig.~\ref{fig:mgbfehbe_sig}. Galaxies outside one $\sigma$ about the
mean of the age gradients' distribution (see Fig.~\ref{fig:gradhisto})
are plotted with smaller symbols.  The typical error is given in the
lower right-hand corner. {\em Right panel:} Residuals from the
correlation shown in the left panel by the solid line as a function of
the age gradient.
\label{fig:gradsigz} 
}
\end{figure*}

While age and the \aFe\ ratio indicate zero gradients, we do find a
significant negative median metallicity gradient (panel b in
Fig.~\ref{fig:gradhisto}).  
%%%new
%{\bf 
The distribution exhibits a well-defined peak at $\triangle
[\ZH]/\triangle \log a_e = -0.16$. The rms of this peak is 0.12~dex
and fully consistent with the median error in the metallicity gradient
($\sim 0.1$~dex, see Table~\ref{tab:ages_gradients}). More than 97 per
cent of the data points fall inside $2\sigma$ around the peak of the
distribution.
%}  
%%%new
The median metallicity gradient found here is in agreement, although a
bit flatter than previous determinations (e.g., Davies et al.\
1993). It has already been discussed extensively in the literature
(e.g., Peletier et al.\ 1990; see also references in the Introduction)
that galaxy formation in a simple dissipative collapse (e.g., Larson
1974) would produce average metallicity gradients of about 0.5 dex per
decade (Carlberg 1984), well in excess of the slope derived
here. Hence some kind of mixing process, for instance induced by
galaxy mergers, are required to flatten out the metallicity gradient
(White 1980).

The total metallicity of a stellar population only depends on the
efficiency of star formation, i.e.\ the {\em fraction} of gas turned
into stars (e.g., Tinsley 1980). It seems therefore natural to assume
that a relationship between metallicity and escape velocity produces
the metallicity gradient (Franx \& Illingworth 1990; Davies et al.\
1993; Martinelli et al.\ 1998).  In the galaxies' center, where the
gravitational potential is highest, star formation and hence the
enrichment of metals is more effective than in outer regions with
lower gravitational potential.  

We find empirical support for this picture, as our data indicate a
correlation between the metallicity gradient and the {\em gradient} in
velocity dispersion (left panel in Fig.~\ref{fig:gradsigz}).  Galaxies
with steeper metallicity gradients have also steeper gradients in
velocity dispersion.  Note that the determination of the metallicity
gradient is affected by uncertainties in the \Hb\ index gradient and
hence the age gradient, due to correlated errors caused by the
age-metallicity degeneracy. In other words, an erroneously steep
negative age gradient causes a too flat or even positive metallicity
gradient and vice versa.  Therefore we determine the linear fit (solid
line in left panel of Fig.~\ref{fig:gradsigz}) for those objects that
are inside $1\sigma$ about the mean of the age gradients' distribution
(see Fig.~\ref{fig:gradhisto}). The galaxies outside one $\sigma$,
hence objects with unusually steep positive or negative age gradients,
are shown as smaller symbols. It can be seen that most outliers fall
in that latter category. The right panel of Fig.~\ref{fig:gradsigz}
illustrates this effect esplicitly, by showing the residuals from the
correlation between the gradients in metallicity and velocity
dispersion (solid line) as a function of the age gradient.  Objects
with age gradients close to the median value follow the correlation
surprisingly well, given the large error bars.

%  As \Hb\ as
%an age indicator can be hampered by horizontal branch morphology
%and/or filling by emission lines, the metallicity gradients of the
%objects with extreme age gradients (most deviating from the median)
%are reliable. For those objects, whose age gradients do not
%significantly deviate from the median value, instead, the correlation
%between the metallicity gradient and the {\em gradient} in velocity
%dispersion is surprisingly well defined, in particular given the large
%error bars.

Finally we note that we did not find any significant evidence for
correlations of the stellar population gradients with central velocity
dispersion or other central stellar population properties, as well as
other galaxy properties like galaxy type, velocity dispersion, and
environmental density in terms of radial distance from the cluster
center.

\subsubsection{Global vs.\ intrinsic Mg-$\sigma$ relations}
%%%new
%{\bf 
In Sect.~\ref{agemetovercentral} we have shown that the
variations of the \aFe\ ratio drive 30\% of the global Mg-$\sigma$
relation. 
% }
%%%
The intrinsic Mg-$\sigma$ relation, i.e.\ {\em
local} velocity dispersion versus {\em local} Mg line-strength inside
a galaxy, instead, must be due to metallicity variations alone,
because of the absence of a significant gradient in \aFe\ and age. In
other words, the global and intrinsic Mg-$\sigma$ relations do not
have a common origin. Somewhat amazingly, however, the slopes of the
global and local relations are not very different.

\section{Conclusions}
\label{conclusion}
From high S/N spectra of 35 early type galaxies in the Coma cluster we
have measured central values and gradients within one effective radius
of velocity dispersion and the line indices \Hb, \Mgb, Fe5270, and
Fe5335.  With the new \aFe\ sensitive stellar population models of
TMB, we have determined central values and gradients of the stellar
population parameters age, total metallicity, and \aFe\ element ratio.

\paragraph{Central values} We find a very clear age dichotomy among the
population of S0 galaxies in the Coma cluster. The first class is
dominated by old ($\sim 10$~Gyr) stellar populations and is
practically indistinguishable from the elliptical galaxy
population. The lenticulars of the second class, that also includes
the two cD galaxies, have very young ($\sim 2$~Gyr) average ages,
hence are dominated by recent major star formation episodes. They may
be the descendants of typical star-forming spirals whose star
formation has been stopped due to the dense cluster environment
(Poggianti et al.\ 2001a).

The element ratio \aFe\ correlates with central velocity
dispersion, and drives 30\% of the Mg-$\sigma$
relation. We find the tendency that older galaxies have higher \aFe\
ratios (see also Thomas et al.\ 2002), which implies that the most
massive galaxies in the Coma cluster had not only the shortest star
formation timescales (because of the \aFe\ ratio; e.g., Matteucci
1994; Thomas et al.\ 1999) but were also the first to form.

Interestingly, none of the indices or the stellar population
parameters depend on the density profile of the Coma cluster, which
spans a range of about 3 dex. We conclude that environmental effects
did not significantly influence on the evolution of the early-type
galaxies in the Coma cluster.

\paragraph{Gradients} 
The most striking result from studying the stellar
population gradients 
%%%new
%{\bf 
(within $1R_e$)
%}
%%%
of our sample is the absence of a gradient in both age and \aFe\
ratio.  The first direct important inference is that the existence of
\aFe\ enhanced stellar populations in early-type galaxies is not
restricted to their centers but it is a global phenomenon. The star
formation timescales do not vary significantly inside a galaxy as a
function of radius. The local potential at a given radius does not
affect the formation timescales.

It does affect, instead, the fraction of gas lost in a galactic wind,
as we detect a significant negative metallicity gradient. At larger
radii, the potential is shallower, i.e.\ the escape velocity is lower,
so that star formation is less complete and the metallicity is lower
than in the center (e.g., Davies et al.\ 1993). This conclusion is
further supported by the fact that we find a relationship between the
metallicity gradient and the gradient in velocity dispersion.

Finally, the presence of a metallicity gradient and the absence of
gradients in age and \aFe\ ratio show that the intrinsic Mg-$\sigma$
relation, i.e.\ {\em local} velocity dispersion versus {\em local} Mg
line-strength inside a galaxy, is driven by metallicity alone. 
%%%new
%{\bf 
We have shown that the global relation, instead, is driven by
both metallicity and \aFe\ ratio, the former contributing roughly 70
the latter 30 per cent.
%}
%%%new
Hence, the global and the intrinsic Mg-$\sigma$ relations have
different origins, the former being at least partially produced by
formation timescales (\aFe\ ratio), the latter by the local
gravitational potential (metallicity) alone.

\begin{acknowledgements} 
%%%new
%{\bf 
We thank the referee Bianca Poggianti for very helpful comments and
Scott Trager for the many interesting discussions. 
% }
%%%
This work was supported by the Deutsche Forschungsgemeinschaft via 
project Be 1091/6 and SFB 375.
\end{acknowledgements}

\newpage
\appendix
\section{Line indices}
\begin{table*}[!ht]
\caption
{The central values (averaged over 0.1 $a_e$) of the indices
H$_{\beta}$, Mg$_b$, Fe$_{5270}$, Fe$_{5335}$ and of the velocity
dispersion $\sigma$.}
\label{tab:indices_zero}
\centerline{
\begin{tabular}{|r|c|c|c|c|c|c|c|c|c|c|c|c|}
\hline  
Nr.&GMP & Type & \Hb&d\Hb&\Mgb&d\Mgb&Fe5270
&dFe5270&Fe5335&dFe5335&$\langle\sigma\rangle$ &  d$\langle\sigma\rangle$\\
&Nr. & & $[$\AA$]$&$[$\AA$]$& $[$\AA$]$&$[$\AA$]$&$[$\AA$]$&$[$\AA$]$&$[$\AA$]$&$[$\AA$]$ &
$[$km/s$]$&$[$km/s$]$\\
\hline
 1 & 3329 &    D & 2.13 & 0.03  & 4.91 & 0.03 & 2.82 & 0.04 &  2.93 & 0.04 & 256.5 & 1.5   \\
 2 & 2921 &    D & 1.94 & 0.07  &  5.45 & 0.08 & 2.93 & 0.08 &  3.15 & 0.09 & 380.6 & 5.1   \\
 3 & 4928 &  E/S0 & 1.41 & 0.10  & 4.90 &  0.11 & 2.83 & 0.12 &  2.68 & 0.13  & 272.3 & 4.7   \\
 4 & 4822 &    E & 1.61 & 0.10  & 4.57 & 0.11 & 2.91 & 0.12 &  2.94 & 0.14       &   262.0 & 3.2   \\
 5 & 1750 &    E & 1.55 & 0.13  & 5.17 & 0.13 &  3.08 & 0.15 &  2.47 & 0.17     & 300.5 & 2.7   \\
 6 & 2390 &    E & 1.43 & 0.13  & 5.33 & 0.14 & 2.93 &  0.15 &  2.57 & 0.17     & 273.8 & 2.7   \\
 7 & 2795 &   S0 & 1.60 & 0.09  & 4.33 & 0.096 & 2.90 &  0.104 &  2.30 & 0.12    & 246.8 & 2.2   \\
 8 & 3792 &    E & 1.39 &  0.13  & 5.39 & 0.14 & 2.99 & 0.15 &  2.72 &  0.17     & 284.7 &  4.9  \\
 9 & 2629 &   S0 & 1.72 & 0.13  &  4.32 & 0.13 & 2.78 & 0.12 &  2.72 & 0.13     & 191.9 & 2.0   \\
10 & 3561 &   S0 & 2.26 &  0.14  &  4.48 &  0.12 & 2.96 & 0.16 &  2.99 & 0.17     &   283.0 & 4.3   \\
11 & 2000 &    E & 1.70 & 0.12  & 4.43 &  0.13 & 2.87 & 0.15 &  2.50 & 0.16     & 199.1 &  2.2  \\
12 & 2413 &   S0 & 1.77 & 0.11  & 4.11 & 0.09 & 2.85 & 0.14 &  2.35 & 0.14    & 211.3 & 2.0   \\
13 & 4829 & E/S0 & 1.63 & 0.10  & 4.94 & 0.12 & 3.01 & 0.12 &  2.81 & 0.13     & 245.5 & 2.9   \\
14 & 3510 &    E & 1.41 & 0.12  & 4.83 &  0.13 & 2.98 & 0.15 &  2.82 & 0.16     & 201.5 & 2.8   \\
15 & 2417 & S0/E & 1.61 & 0.17  & 4.55 & 0.18 & 2.82 & 0.20 &  2.39 & 0.22     & 224.4 & 3.8   \\
16 & 2440 &    E & 1.46 & 0.15  & 4.70 & 0.17 &  2.95 & 0.19 &  2.58 & 0.21       & 209.4 &  3.7  \\
17 & 3414 &   S0 & 1.61 & 0.22  & 4.36 & 0.23 & 2.82 & 0.27 &  2.41 & 0.29     & 168.7 & 9.5   \\
18 & 4315 & E/S0 & 1.57 & 0.15  & 4.39 & 0.16 & 2.67 & 0.19 &  2.50 & 0.21     & 190.2 &   4.1 \\
19 & 3073 &   S0 & 1.58 & 0.14  & 4.44 &  0.15 & 3.12 & 0.16 &  2.60 &  0.18     & 170.4 &  2.4  \\
20 & 1853 &   S0 & 1.65 & 0.22  & 4.33 & 0.24 & 2.91 & 0.25 &  2.75 & 0.28     & 201.8 & 3.8   \\
21 & 3201 &    E & 2.29 & 0.42  & 4.12 & 0.44 & 2.53 & 0.48 &  1.88 & 0.56     & 181.7 &   5.7 \\
22 & 3661 &   S0 & 2.25 & 0.16  & 4.20 & 0.17 & 2.79 & 0.18 &  2.86 & 0.20     & 177.8 & 7.3   \\
23 & 4679 &   S0 & 2.37 & 0.12  & 3.42 & 0.13 & 2.45 & 0.14 &  2.62 & 0.16     & 82.0 & 2.2   \\
24 & 3352 & E/S0 & 2.05 & 0.17  & 4.00 & 0.18 & 2.80 & 0.19 &  2.73 &  0.22     & 186.7 &   6.7 \\
25 & 2535 &   S0 & 2.22 & 0.08  & 3.93 & 0.09 & 2.91 & 0.09 &  2.88 & 0.10  & 124.4 & 3.0   \\
26 & 3958 &    E & 1.29 & 0.14  & 4.18 & 0.15 & 2.29 &  0.16 &  1.59 & 0.18     & 152.9 &   2.1 \\
27 & 2776 & S0/E & 2.00 & 0.2   & 4.21 & 0.22 & 2.69 & 0.24 &  1.90 & 0.27     & 132.6 &  2.8  \\
28 & 0144 &    E & 1.76 & 0.04  & 4.53 & 0.04 & 2.89 & 0.05 &  2.98 & 0.05 & 217.1 & 1.4   \\
29 & 0282 &    E & 1.71 & 0.07  & 4.75 & 0.08 & 2.88 & 0.09 &  2.51 & 0.10  & 290.6 & 2.3   \\
30 & 0756 &   S0 & 2.15 & 0.06  & 3.83 & 0.07 & 2.70 & 0.08 & 2.55 & 0.09    & 178.3 & 0.7 \\
31 & 1176 &   S0 & 2.13 & 0.07  & 3.87 & 0.08 & 2.80 & 0.09 &  2.55 & 0.10 &   189.0 & 0.8 \\
32 & 1990 & E/S0 & 1.40 & 0.04  & 4.78 & 0.12 & 2.83 & 0.13 &  2.19 & 0.15     & 275.1 & 3.7 \\
33 & 5279 &    E & 1.54 & 0.093  & 4.89 & 0.10 & 2.91 & 0.12 &  2.70 & 0.13    & 269.1 & 2.7 \\
34 & 5568 &   S0 & 1.62 & 0.066  &  4.82 & 0.07 & 2.80 & 0.08 &  2.63 & 0.09  & 226.9 & 1.8\\ 
35 & 5975 &    E & 1.81 & 0.18  & 4.39 & 0.20 & 2.88 &  0.23 &  2.68 & 0.25     & 198.4 & 5.4 \\
\hline                                     
\end{tabular}
}
\end{table*}

\begin{table*}[!ht]
\caption
{The fitted logarithmic
gradients inside $a<1 a_e$ of the indices 
H$_{\beta}$, Mg$_b$, Fe$_{5270}$,
Fe$_{5335}$ and the velocity dispersion $\sigma$ .} 
\label{tab:indices_gradients}
\centerline{
\begin{tabular}{|c|r|c|r|c|r|c|r|c|r|c|}
\hline  
GMP Nr.& $\triangle$H$_{\beta}$&d$\triangle$H$_{\beta}$ &
$\triangle$Mg$_b$ & d$\triangle$Mg$_b$ & $\triangle$Fe$_{5270}$ &
d$\triangle$Fe$_{5270}$ & $\triangle$Fe$_{5335}$& d$\triangle$Fe$_{5335}$ &
$\triangle \sigma$& d$\triangle \sigma$ \\
\hline
  3329 & -0.006 &  0.010 & -0.010 &  0.005 & -0.035 &  0.009 &   -0.006 &  0.009 & -0.025 &  0.009 \\ 
  2921 & -0.026 &  0.023 & -0.021 &  0.009 & -0.036 &  0.017 &   -0.045 &  0.018 & -0.065 &  0.015 \\ 
  4928 &  0.029 &  0.037 & -0.046 &  0.013 & -0.026 &  0.024 &    0.064 &  0.023 & -0.004 &  0.024 \\ 
  4822 &  0.029 &  0.037 & -0.046 &  0.013 & -0.026 &  0.024 &    0.064 &  0.023 & -0.023 &  0.020 \\ 
  1750 & -0.050 &  0.048 & -0.062 &  0.016 & -0.046 &  0.028 &    0.025 &  0.036 & -0.077 &  0.012 \\ 
  2390 & -0.044 &  0.057 & -0.045 &  0.015 & -0.019 &  0.027 &   -0.022 &  0.035 & -0.034 &  0.013 \\ 
  2795 &  0.042 &  0.015 & -0.031 &  0.007 & -0.002 &  0.011 &    0.004 &  0.018 & -0.128 &  0.008 \\ 
  3792 &  0.045 &  0.044 & -0.042 &  0.014 & -0.034 &  0.027 &    0.007 &  0.033 & -0.014 &  0.021 \\ 
  2629 &  0.110 &  0.027 &  0.001 &  0.013 & -0.002 &  0.019 &    0.082 &  0.020 & -0.005 &  0.012 \\ 
  3561 & -0.010 &  0.034 & -0.038 &  0.013 & -0.017 &  0.027 &   -0.042 &  0.029 & -0.103 &  0.014 \\ 
  2000 & -0.001 &  0.032 & -0.041 &  0.014 & -0.026 &  0.024 &   -0.039 &  0.032 & -0.053 &  0.016 \\ 
  2413 & -0.011 &  0.025 & -0.018 &  0.008 &  0.016 &  0.018 &    0.006 &  0.026 & -0.059 &  0.008 \\ 
  4829 &  0.002 &  0.029 & -0.038 &  0.010 & -0.037 &  0.018 &   -0.056 &  0.021 & -0.075 &  0.011 \\ 
  3510 &  0.038 &  0.031 & -0.045 &  0.011 & -0.050 &  0.021 &   -0.056 &  0.025 & -0.050 &  0.017 \\ 
  2417 &  0.076 &  0.034 &  0.016 &  0.015 &  0.015 &  0.022 &   -0.031 &  0.030 & -0.055 &  0.013 \\ 
  2440 &  0.064 &  0.034 & -0.050 &  0.014 & -0.046 &  0.024 &   -0.005 &  0.028 & -0.169 &  0.018 \\ 
  3414 &  0.051 &  0.049 & -0.034 &  0.021 &  0.018 &  0.034 &   -0.063 &  0.048 & -0.088 &  0.021 \\ 
  4315 &  0.008 &  0.035 & -0.036 &  0.013 & -0.024 &  0.024 &   -0.056 &  0.029 & -0.106 &  0.015 \\ 
  3073 &  0.037 &  0.039 & -0.003 &  0.015 &  0.009 &  0.024 &   -0.008 &  0.032 & -0.067 &  0.019 \\ 
  1853 &  0.043 &  0.060 & -0.054 &  0.028 & -0.044 &  0.045 &   -0.058 &  0.055 & -0.151 &  0.021 \\ 
  3201 & -0.136 &  0.085 &  0.043 &  0.034 &  0.011 &  0.058 &    0.165 &  0.070 &  0.022 &  0.031 \\ 
  3661 &  0.150 &  0.022 & -0.037 &  0.016 & -0.069 &  0.027 &   -0.047 &  0.027 & -0.101 &  0.020 \\ 
  4679 &  0.003 &  0.031 &  0.040 &  0.022 & -0.000 &  0.034 &    0.045 &  0.033 & -0.005 &  0.035 \\ 
  3352 &  0.076 &  0.022 & -0.029 &  0.014 & -0.026 &  0.021 &   -0.094 &  0.029 & -0.108 &  0.027 \\ 
  2535 & -0.027 &  0.022 & -0.051 &  0.013 & -0.069 &  0.019 &   -0.044 &  0.021 & -0.056 &  0.017 \\ 
  3958 &  0.005 &  0.052 & -0.027 &  0.017 & -0.028 &  0.034 &   -0.046 &  0.058 & -0.037 &  0.021 \\ 
  2776 &  0.034 &  0.043 & -0.019 &  0.024 & -0.058 &  0.042 &    0.053 &  0.058 & -0.015 &  0.015 \\ 
  0144 &  0.038 &  0.015 & -0.055 &  0.007 & -0.019 &  0.012 &   -0.031 &  0.013 & -0.062 &  0.006 \\ 
  0282 &  0.022 &  0.022 & -0.056 &  0.010 & -0.042 &  0.018 &   -0.003 &  0.022 & -0.069 &  0.006 \\ 
  0756 &  0.013 &  0.017 & -0.071 &  0.011 & -0.054 &  0.019 &   -0.053 &  0.022 & -0.124 &  0.006 \\ 
  1176 &  0.012 &  0.014 & -0.029 &  0.009 & -0.028 &  0.014 &   -0.019 &  0.016 & -0.084 &  0.004 \\ 
  1990 & -0.002 &  0.040 &  0.015 &  0.014 & -0.036 &  0.024 &    0.087 &  0.030 & -0.095 &  0.009 \\ 
  5279 &  0.058 &  0.032 & -0.068 &  0.013 & -0.056 &  0.025 &    0.012 &  0.027 & -0.063 &  0.011 \\ 
  5568 &  0.005 &  0.039 & -0.035 &  0.014 & -0.046 &  0.029 &   -0.012 &  0.032 & -0.047 &  0.012 \\ 
  5975 &  0.017 &  0.040 & -0.087 &  0.020 & -0.029 &  0.033 &   -0.023 &  0.037 & -0.078 &  0.016 \\ 
\hline                                     
\end{tabular}
}
\end{table*}

\begin{table*}[!ht]
\caption
{Logarithmic fit values at $ 1 a_e$ of the 
indices H$_{\beta}$, Mg$_b$, Fe$_{5270}$,
Fe$_{5335}$ and the velocity dispersion $\sigma$ .} 
\label{tab:indices_re_fit}
\centerline{
%\begin{tabular}{|c|p{1.0cm}|p{1.15cm}|p{1.15cm}|p{1.3cm}|p{1.4cm}|p{1.55cm}|p{1.4cm}|p{1.55cm}|p{0.9cm}|p{0.95cm}|}
\begin{tabular}{|c|c|c|c|c|c|c|c|c|c|c|}
\hline  
%GMP Nr.&$\log_{a_e}$H$_{\beta}$&d$\log_{a_e}$H$_{\beta}$&$\log_{a_e}$Mg$_b$&d$\log_{a_e}$Mg$_b$&$\log_{a_e}$Fe$_{5270}$&d$\log_{a_e}$Fe$_{5270}$&$\log_{a_e}$Fe$_{5335}$&d$\log_{a_e}$Fe$_{5335}$&$\log_{a_e} \sigma$&d$\log_{a_e} \sigma$ \\
GMP Nr.&$\log_{a_e}$&d$\log_{a_e}$&$\log_{a_e}$&d$\log_{a_e}$&$\log_{a_e}$&d$\log_{a_e}$&$\log_{a_e}$&d$\log_{a_e}$&$\log_{a_e}$&d$\log_{a_e}$ \\
       & H$_{\beta}$&H$_{\beta}$  &    Mg$_b$  &Mg$_b$       &Fe$_{5270}$ &Fe$_{5270}$  &Fe$_{5335}$ &Fe$_{5335}$  &   $\sigma$ & $\sigma$\\
\hline
  3329 & 0.325 & 0.016 & 0.678 & 0.008 & 0.397 & 0.014    & 0.474 & 0.014 & 2.397 & 0.015 \\ 
  2921 & 0.246 & 0.030 & 0.708 & 0.012 & 0.409 & 0.022    & 0.426 & 0.024 & 2.477 & 0.020 \\ 
  4928 & 0.205 & 0.041 & 0.625 & 0.016 & 0.412 & 0.028    & 0.512 & 0.025 & 2.431 & 0.033 \\ 
  4822 & 0.196 & 0.032 & 0.639 & 0.012 & 0.420 & 0.022    & 0.493 & 0.020 & 2.387 & 0.024 \\ 
  1750 & 0.114 & 0.049 & 0.636 & 0.016 & 0.433 & 0.028    & 0.446 & 0.033 & 2.379 & 0.012 \\ 
  2390 & 0.098 & 0.063 & 0.674 & 0.016 & 0.435 & 0.029    & 0.372 & 0.038 & 2.384 & 0.014 \\ 
  2795 & 0.290 & 0.016 & 0.594 & 0.008 & 0.454 & 0.013    & 0.384 & 0.018 & 2.225 & 0.007 \\ 
  3792 & 0.198 & 0.033 & 0.680 & 0.011 & 0.445 & 0.022    & 0.461 & 0.025 & 2.434 & 0.020 \\ 
  2629 & 0.400 & 0.028 & 0.632 & 0.015 & 0.444 & 0.022    & 0.566 & 0.020 & 2.278 & 0.018 \\ 
  3561 & 0.319 & 0.028 & 0.612 & 0.010 & 0.446 & 0.021    & 0.413 & 0.023 & 2.340 & 0.010 \\ 
  2000 & 0.232 & 0.028 & 0.599 & 0.013 & 0.425 & 0.022    & 0.357 & 0.029 & 2.229 & 0.016 \\ 
  2413 & 0.228 & 0.028 & 0.601 & 0.010 & 0.493 & 0.018    & 0.404 & 0.026 & 2.237 & 0.010 \\ 
  4829 & 0.228 & 0.022 & 0.648 & 0.008 & 0.432 & 0.015    & 0.382 & 0.019 & 2.298 & 0.009 \\ 
  3510 & 0.202 & 0.026 & 0.625 & 0.010 & 0.406 & 0.019    & 0.370 & 0.024 & 2.245 & 0.015 \\ 
  2417 & 0.269 & 0.023 & 0.642 & 0.010 & 0.460 & 0.017    & 0.325 & 0.026 & 2.255 & 0.014 \\ 
  2440 & 0.240 & 0.021 & 0.616 & 0.009 & 0.420 & 0.016    & 0.408 & 0.018 & 2.140 & 0.014 \\ 
  3414 & 0.258 & 0.033 & 0.611 & 0.015 & 0.461 & 0.023    & 0.326 & 0.034 & 2.116 & 0.024 \\ 
  4315 & 0.199 & 0.033 & 0.595 & 0.011 & 0.392 & 0.020    & 0.311 & 0.025 & 2.134 & 0.014 \\ 
  3073 & 0.240 & 0.031 & 0.643 & 0.012 & 0.490 & 0.019    & 0.403 & 0.026 & 2.159 & 0.017 \\ 
  1853 & 0.263 & 0.044 & 0.582 & 0.023 & 0.423 & 0.036    & 0.394 & 0.045 & 2.132 & 0.016 \\ 
  3201 & 0.164 & 0.075 & 0.637 & 0.023 & 0.453 & 0.040    & 0.499 & 0.040 & 2.276 & 0.027 \\ 
  3661 & 0.477 & 0.013 & 0.585 & 0.011 & 0.383 & 0.019    & 0.419 & 0.019 & 2.137 & 0.016 \\ 
  4679 & 0.383 & 0.029 & 0.590 & 0.020 & 0.391 & 0.033    & 0.481 & 0.031 & 1.903 & 0.036 \\ 
  3352 & 0.356 & 0.011 & 0.586 & 0.007 & 0.431 & 0.011    & 0.379 & 0.016 & 2.208 & 0.017 \\ 
  2535 & 0.320 & 0.017 & 0.542 & 0.010 & 0.394 & 0.015    & 0.413 & 0.017 & 2.031 & 0.016 \\ 
  3958 & 0.118 & 0.031 & 0.599 & 0.011 & 0.337 & 0.021    & 0.161 & 0.036 & 2.149 & 0.014 \\ 
  2776 & 0.321 & 0.026 & 0.603 & 0.015 & 0.377 & 0.027    & 0.329 & 0.034 & 2.084 & 0.014 \\ 
  0144 & 0.283 & 0.013 & 0.597 & 0.006 & 0.437 & 0.010    & 0.432 & 0.011 & 2.283 & 0.005 \\ 
  0282 & 0.254 & 0.017 & 0.610 & 0.008 & 0.412 & 0.014    & 0.400 & 0.016 & 2.366 & 0.006 \\ 
  0756 & 0.348 & 0.013 & 0.500 & 0.009 & 0.376 & 0.015    & 0.352 & 0.018 & 2.084 & 0.006 \\ 
  1176 & 0.351 & 0.011 & 0.548 & 0.007 & 0.413 & 0.012    & 0.382 & 0.014 & 2.146 & 0.004 \\ 
  1990 & 0.127 & 0.023 & 0.668 & 0.007 & 0.412 & 0.016    & 0.433 & 0.018 & 2.267 & 0.011 \\ 
  5279 & 0.245 & 0.023 & 0.616 & 0.010 & 0.405 & 0.019    & 0.430 & 0.020 & 2.340 & 0.012 \\ 
  5568 & 0.218 & 0.052 & 0.634 & 0.019 & 0.379 & 0.039    & 0.406 & 0.042 & 2.296 & 0.020 \\ 
  5975 & 0.269 & 0.022 & 0.572 & 0.012 & 0.432 & 0.018    & 0.420 & 0.021 & 2.215 & 0.013 \\ 
\hline                                     
\end{tabular}
}
\end{table*}

%\clearpage
\clearpage
\section{Stellar parameters}
\begin{table*}[!ht]
\caption
{The central ages, metallicities ($[Z/$H$]$) and the $[\aFe]$ ratios
derived from the indices listed in Tab.~\ref{tab:indices_zero}
combined with the models from TMB (see Fig.~\ref{fig:mgfehb} as well
as Sect.~\ref{model}).}
\label{tab:ages_zero}
%\begin{tabular}{|c|r|r|p{0.7cm}|p{0.7cm}|p{0.7cm}|p{0.9cm}|}
\begin{tabular}{|c|r|r|c|c|c|c|}
\hline 
GMP Nr.& age & dage & $[Z/$H$]$ & d$[Z/$H$]$ & 
$[\aFe]$ & d$[\aFe]$\\
& [Gyr] & [Gyr] & & & & \\
\hline
 3329 & 2.2 &  0.1 &  0.77 &  0.04 &  0.37 &  0.01 \\
 2921 & 2.6 &  0.2 &  0.75 &  0.05 &  0.42 &  0.02 \\  
 4928 &  14.5  &  1.4  &  0.15  &  0.05 &  0.28  &  0.03     \\
 4822 &  11.2  &  1.3  &  0.21  &  0.06 &  0.16  &   0.03  \\
 1750 &  11.3  &  1.7  &  0.28  &   0.08 &  0.33  &  0.03    \\
 2390 &  13.2  &   2.0  &  0.26  &  0.08 &  0.36  &  0.03    \\
 2795 &   9.6  &  1.3  &  0.11  &   0.06 &  0.25  &  0.03     \\
 3792 &  13.4  &   2.1  &  0.31  &  0.07 &  0.33  &  0.03    \\
 2629 &   8.7  &  1.5  &  0.12  &  0.07 &  0.14  &  0.03    \\
 3561 &   2.0  &  0.15 &  0.854 &  0.163 &   0.287 &  0.04  \\
 2000 &   8.6  &  1.3  &  0.18  &   0.06 &  0.24  &  0.03    \\
 2413 &   7.3  &  1.6  &  0.10  &    0.06 &  0.21  &   0.03     \\
 4829 &   7.9  &  1.9  &  0.37  &  0.06 &  0.27  &   0.03    \\
 3510 &  14.2  &  1.6  &  0.19  &   0.06 &  0.22  &   0.03    \\
 2417 &  11.5  &   2.4  &  0.10  &  0.10 &  0.27  &   0.05    \\
 2440 &  13.5  &   2.1  &  0.13  &   0.08 &  0.24  &  0.04  \\
 3414 &  11.2  &   2.7  &  0.07  &   0.12 &  0.23  &   0.07    \\
 4315 &  12.2  &   1.7  &  0.04  &   0.08 &  0.24  &   0.05    \\
 3073 &  10.5  &   1.5  &  0.19  &   0.07 &  0.17  &  0.04  \\
 1853 &  10.1  &   2.9  &  0.16  &  0.12 &  0.15  &  0.06    \\
 3201    &   2.5 &   1.1 &  0.22 &  0.24 &  0.45 &  0.13 \\
 3661    &   2.0 &   0.2 &  0.59 &  0.14 & 0.25 &  0.05 \\
 4679 &   2.1  &  0.3  &  0.20  &   0.06 &  0.14  &  0.04    \\
 3352 &   3.4  &   0.5  &  0.32  &  0.08 &  0.20  &   0.04    \\
 2535 &   2.3  &   0.5  &  0.41  &  0.09 &  0.14  &  0.03     \\
 3958 &   $-\ $ & $-\ $ &  $  -$ & $  -$ &  $  -$ & $  -$\\ 
 2776    &  11.8  & 2.3 &  -0.10 &  0.06  & 0.28 &  0.09  \\  
 0144 &   5.8  &    0.5  &  0.34  &   0.02 &  0.20  &   0.01    \\
 0282 &   7.7  &  0.8  &  0.26  &  0.04 &  0.30  &    0.02    \\
 0756 &   3.1  &   0.2  &  0.24  &  0.03 &  0.20  &  0.02  \\
 1176 &   3.3  &  0.4  &  0.26  &   0.04 &  0.19  &   0.02    \\
 1990 &  17.0  &   1.2  &$-0.01$ &   0.05 &  0.33  &  0.04    \\
 5279 &  10.9  &  0.8  &  0.25  &  0.05 &  0.27  &  0.03     \\
 5568 &   9.6  &   0.6  &  0.25  &  0.04 &  0.30  &  0.02    \\ 
 5975 &   5.7  &  1.2  &  0.27  &  0.09 &  0.22  &  0.05    \\
\hline                                     
\end{tabular}
\end{table*}

\begin{table*}[!ht]
\caption
{Gradients of the ages, metallicities ($[Z/$H$]$) and $[\aFe]$) ratios
derived from the fitted gradients values at $1 a_e$ listed in
Tab.~\ref{tab:indices_gradients} and Tab.~\ref{tab:indices_re_fit}
combined with the models from TMB (see Sect.~\ref{agemetovergradients}
for details.)}
\label{tab:ages_gradients}
%\begin{tabular}{|c|p{0.6cm}|p{0.7cm}|p{0.85cm}|p{1.05cm}|p{0.9cm}|p{1.15cm}|}
\begin{tabular}{|c|c|c|c|c|c|c|}
\hline 
GMP& $\triangle$age & d$\triangle$age & $\triangle [Z/$H$]$ &
d$\triangle [Z/$H$]$ & 
$\triangle [\aFe]$ & d$\triangle [\aFe]$\\
Nr.& [Gyr] & [Gyr] & & & & \\
\hline
 3329 &   $ 0.1$ & $0.3$ &  $-0.19$ & $0.07$ &  $ 0.02$ & $0.02$\\        
 2921 &   $ 4.2$ & $1.3$ &  $-0.29$ & $0.12$ &  $-0.02$ & $0.03$\\        
 4928 &   $ 0.2$ & $3.4$ &  $-0.06$ & $0.10$ &  $-0.16$ & $0.04$\\        
 4822 &   $ 2.4$ & $3.2$ &  $-0.14$ & $0.09$ &  $-0.01$ & $0.04$\\        
 1750 &   $ 4.8$ & $5.2$ &  $-0.25$ & $0.13$ &  $-0.09$ & $0.07$\\        
 2390 &   $ 6.5$ & $5.8$ &  $-0.26$ & $0.14$ &  $ 0.02$ & $0.06$\\        
 2795 &   $-2.5$ & $1.4$ &  $ 0.01$ & $0.05$ &  $-0.05$ & $0.02$\\        
 3792 &   $-2.6$ & $4.8$ &  $-0.02$ & $0.12$ &  $-0.01$ & $0.06$\\        
 2629 &   $-0.7$ & $3.2$ &  $ 0.04$ & $0.29$ &  $ 0.03$ & $0.08$\\ 
 3561 &   $-0.3$ & $0.8$ &  $-0.28$ & $0.19$ &  $ 0.04$ & $0.06$\\ 
 2000 &   $ 0.6$ & $2.5$ &  $-0.11$ & $0.10$ &  $-0.01$ & $0.06$\\        
 2413 &   $ 1.8$ & $2.2$ &  $-0.07$ & $0.07$ &  $-0.10$ & $0.04$\\        
 4829 &   $-0.2$ & $2.5$ &  $-0.17$ & $0.08$ &  $ 0.03$ & $0.04$\\        
 3510 &   $ 0.3$ & $2.5$ &  $-0.19$ & $0.08$ &  $ 0.04$ & $0.04$\\ 
 2417 &   $-7.3$ & $2.9$ &  $ 0.21$ & $0.07$ &  $ 0.09$ & $0.05$\\
 2440 &   $-5.3$ & $2.9$ &  $ 0.02$ & $0.10$ &  $-0.02$ & $0.04$\\         
 3414 &   $ 2.8$ & $4.4$ &  $-0.13$ & $0.14$ &  $-0.03$ & $0.09$\\        
 4315 &   $ 0.1$ & $3.0$ &  $-0.10$ & $0.07$ &  $ 0.04$ & $0.04$\\
 3073 &   $-3.5$ & $3.7$ &  $ 0.07$ & $0.11$ &  $ 0.04$ & $0.06$\\        
 1853 &   $ 0.4$ & $4.2$ &  $-0.20$ & $0.18$ &  $-0.01$ & $0.09$\\        
 3201 &   $17.9$ & $7.5$ &  $-0.10$ & $0.30$ &  $-0.10$ & $0.14$\\        
 3661 &   $  -$ & $  -$ &  $  -$ & $  -$ &  $  -$ & $  -$\\ 
 4679 &   $  -$ & $  -$ &  $  -$ & $  -$ &  $  -$ & $  -$\\ 
 3352 &   $-0.6$ & $1.1$ &  $-0.17$ & $0.11$ &  $ 0.12$ & $0.04$\\        
 2535 &   $ 0.8$ & $0.8$ &  $-0.23$ & $0.12$ &  $-0.06$ & $0.05$\\        
 3958 &   $  -$ & $  -$ &  $  -$ & $  -$ &  $  -$ & $  -$\\ 
 2776 &   $-0.1$ & $2.5$ &  $ 0.06$ & $0.13$ &  $-0.02$ & $0.09$\\        
 0144 &   $ 0.0$ & $1.2$ &  $-0.18$ & $0.04$ &  $-0.02$ & $0.02$\\        
 0282 &   $ 0.2$ & $2.0$ &  $-0.20$ & $0.07$ &  $-0.06$ & $0.03$\\        
 0756 &   $-0.0$ & $0.6$ &  $-0.20$ & $0.06$ &  $-0.02$ & $0.05$\\        
 1176 &   $-0.2$ & $0.4$ &  $-0.10$ & $0.05$ &  $-0.02$ & $0.03$\\        
 1990 &   $ 1.6$ & $4.0$ &  $-0.08$ & $0.07$ &  $ 0.12$ & $0.06$\\        
 5279 &   $-2.6$ & $3.5$ &  $-0.10$ & $0.11$ &  $-0.05$ & $0.06$\\        
 5568 &   $ 2.1$ & $3.4$ &  $-0.21$ & $0.12$ &  $ 0.00$ & $0.06$\\        
 5975 &   $ 2.9$ & $2.4$ &  $-0.34$ & $0.16$ &  $-0.11$ & $0.07$\\        
\hline                                     
\end{tabular}
\end{table*}

\end{document}